\documentclass[a4paper,11pt]{article}
\pdfoutput=1 %

\usepackage{jheppub}
\usepackage{amsmath,amssymb,graphicx,float,slashed,xcolor,multicol}%,amsthm
\usepackage{newtxmath}
\usepackage{graphicx,tabularx}
\usepackage{url}
\usepackage{footmisc}
\usepackage{amsfonts}
\usepackage{cancel}
\usepackage{color}
\usepackage{multirow}
\usepackage{amssymb}
\usepackage{pifont}
\usepackage{epstopdf}
\usepackage{slashed}
\usepackage{comment}
\usepackage{booktabs}
\usepackage{natbib}
\usepackage{array}
\usepackage{mathrsfs}
\usepackage{subcaption}
\usepackage[toc,page]{appendix}
\usepackage{mathtools}
\usepackage{romannum}
\usepackage{ulem}
\usepackage{bbold}
\usepackage{enumitem}
\usepackage{multirow}

\usepackage{amsmath,amssymb,graphicx,float,slashed,xcolor,multicol}
\usepackage{caption,subcaption}
\usepackage{float}
\usepackage{multirow}
\usepackage{slashed}
\usepackage{upgreek}
\usepackage[toc,page]{appendix}
\usepackage{romannum}
\usepackage{lipsum}
\usepackage{lineno}
\newcommand{\tabincell}[2]{\begin{tabular}{@{}#1@{}}#2\end{tabular}}
\allowdisplaybreaks

\newcounter{magicrownumbers}
\newcommand*{\rom}[1]{\expandafter\@slowromancap\romannumeral #1@}

\def\beq{\begin{equation}}
\def\eeq{\end{equation}}
\title{Damping signatures at JUNO, a medium-baseline reactor neutrino oscillation experiment}

\author[1]{Jun Wang}
\author[1]{Jiajun Liao\footnote{Corresponding author.}}
\author[1]{Wei Wang\footnote{Corresponding author.}}
\author[6]{Angel Abusleme}
\author[45]{Thomas Adam}
\author[66]{Shakeel Ahmad}
\author[66]{Rizwan Ahmed}
\author[55]{Sebastiano Aiello}
\author[66]{Muhammad Akram}
\author[29]{Fengpeng An}
% ...
\affiliation[1]{Sun Yat-Sen University, Guangzhou, China}
% ...
\affiliation[6]{Pontificia Universidad Cat\'{o}lica de Chile, Santiago, Chile}
% ...
\affiliation[29]{East China University of Science and Technology, Shanghai, China}
% ...
\affiliation[45]{IPHC, Universit\'{e} de Strasbourg, CNRS/IN2P3, F-67037 Strasbourg, France}
% ...
\affiliation[55]{INFN Catania and Dipartimento di Fisica e Astronomia dell Universit\`{a} di Catania, Catania, Italy}
% ...
\affiliation[66]{Pakistan Institute of Nuclear Science and Technology, Islamabad, Pakistan}
% ...

\abstract{We study damping signatures at the Jiangmen Underground Neutrino Observatory (JUNO), a medium-baseline reactor neutrino oscillation experiment. These damping signatures are motivated by various new physics models, including quantum decoherence, $\nu_3$ decay, neutrino absorption, and wave packet decoherence. The phenomenological effects of these models can be characterized by exponential damping factors at the probability level. We assess how well JUNO can constrain these damping parameters and how to disentangle these different damping signatures at JUNO. Compared to current experimental limits, JUNO can significantly improve the limits on $\tau_3/m_3$ in the $\nu_3$ decay model, the width of the neutrino wave packet $\sigma_x$, and the intrinsic relative dispersion of neutrino momentum $\sigma_{\rm rel}$.  
}
\begin{document}

\titlepage

\maketitle

%\newpage

%%%%%%%%%%%%%%%%%%%%%%%%%%%

\flushbottom

%%%%%%%%%%%%%%%%%%%%%%%%%%%
\section{Introduction}
\label{sec:intro}
%Neutrino oscillation has been well confirmed in various neutrino experiments using solar, atmospheric, reactor, and accelerator neutrinos~\cite{Tanabashi:2018oca}. 

 Neutrino oscillation was first proposed by Bruno Pontecovero in 1957~\cite{Pontecorvo:1957qd} and was invoked for the solution of atmospheric neutrino anomaly and solar neutrino puzzle. It was experimentally confirmed by the Super-Kamioka Neutrino Detection Experiment (Super-K, SK)~\cite{Super-Kamiokande:1998kpq} in 1998 and the Sudbury Neutrino Observatory (SNO)~\cite{Giganti:2017fhf} in 2002; for further details see Ref.~\cite{Zyla:2020zbs}. Most neutrino oscillation experiments can be well explained in the Standard Model (SM) with three massive neutrinos.
In the standard three-flavor neutrino oscillation framework, the three known neutrino flavor eigenstates
($\nu_e$, $\nu_\mu$, and $\nu_\tau$) can be written as quantum superpositions of three mass
eigenstates ($\nu_1$, $\nu_2$, and $\nu_3$), and the neutrino oscillation probabilities are expressed in terms of
six oscillation parameters: three mixing angles ($\theta_{12}$,
$\theta_{13}$, and $\theta_{23}$), two mass-squared differences ($\Delta
m_{21}^2$ and $\Delta m_{31}^2$), and one Dirac CP phase
($\delta_\text{CP}$). The Majorana CP phases play no role in neutrino oscillations if neutrinos are Majorana particles. Among these six observable oscillation parameters, $\Delta m_{21}^2$, $|\Delta
m_{31}^2|$, $\theta_{12}$, and $\theta_{13}$ have been
well determined to the few-percent level. 
However, the neutrino mass ordering (whether $\Delta m_{31}^2$ is positive or negative), the octant of $\theta_{23}$ (whether $\theta_{23}$ is larger or smaller than 45$^\circ$) and the Dirac CP phase are still open questions. 
At present, the normal mass ordering (NMO) and the second octant of $\theta_{23}$ are both favored by less than $3\sigma$ confidence level (CL)~\cite{NOvA:2017ohq,Super-Kamiokande:2017yvm,Zyla:2020zbs}, and $\delta_\text{CP}$ is in the range of [-3.41, -0.03] for the NMO and [-2.54, -0.32] for the inverted mass ordering (IMO) at the $3\sigma$ CL~\cite{T2K:2019bcf}, respectively.
The main physics goals of next-generation neutrino oscillation experiments, such as the Deep Underground Neutrino Experiment (DUNE)~\cite{Acciarri:2015uup,DUNE:2020jqi}, Hyper-Kamiokande~\cite{Abe:2018uyc} and the Jiangmen Underground Neutrino Observatory (JUNO)~\cite{An:2015jdp,Abusleme:2021zrw}, are to determine the mass ordering with a $3-5\sigma$ CL and to observe CP violation with a $3\sigma$ CL for $\sim75\%$ of $\delta_\text{CP}$ values, etc. To reach these goals, the ability to achieve high-precision measurement of the oscillation spectrum is required for these experiments.
In the meantime, these high-precision experiments will also reach sufficient sensitivity to probe new physics beyond the standard three-neutrino paradigm.

%potentially leading to a different number of neutrinos observed than expected
%potentially leading to fewer neutrinos observed than expected
%Neutrino oscillation is a quantum phenomenon of spontaneous periodic flavor change that occurs when massive neutrinos propagating in space.
The presence of new physics in the neutrino sector would yield corrections to the standard three-flavor neutrino oscillation probabilities, thus leading to modifications to the spectrum measured in high-precision neutrino oscillation experiments.
Among various possible new physics scenarios, a number of them lead to exponential damping in the neutrino oscillation probabilities~\cite{Blennow:2005yk,Blennow:2006hd}, which could yield a different number of neutrinos observed than expected~\cite{Blennow:2006hd, MINOS:2010fgd, Abrahao:2015rba,Choubey:2018cfz, Porto-Silva:2020gma,Ghoshal:2020hyo} or a shift in the best fit values for neutrino oscillation parameters~\cite{Blennow:2005yk, Blennow:2006hd, Fogli:2007tx, MINOS:2010fgd, Abrahao:2015rba, Chan:2015mca,An:2016pvi, Coelho:2017zes,Choubey:2018cfz, deGouvea:2020hfl, Cheng:2020jje}.
%exhibit damping signatures at the probability level~\cite{Blennow:2005yk,Blennow:2006hd}. 
These damping signatures can be treated as secondary effects relative to the standard three-neutrino oscillations in the neutrino flavor transitions. In this work, we present a systematic study of the possible damping effects at the JUNO detector. JUNO is a medium-baseline reactor neutrino experiment with a 20kton liquid scintillator (LS) detector located in a laboratory at 700m underground in Jiangmen, China. The main physics goals of JUNO are to determine the mass ordering and perform high-precision measurements of the neutrino oscillation parameters $\sin^2\theta_{12}$, $\Delta
m_{21}^2$ and $|\Delta m_{ee}^2|$~\cite{An:2015jdp,Abusleme:2021zrw}. 
Also, JUNO is expected to be sensitive to the tiny damping signatures due to its effective energy resolution of 3\% at 1 MeV and the capability of measuring multiple oscillation cycles~\cite{Cheng:2020jje}.

This paper is organized as follows. In Section~\ref{sec:models}, we discuss the damping signatures arising from different new physics models. In Section~\ref{sec:dampingeffects}, we discuss the damping signatures at medium-baseline reactor neutrino experiments. In Section~\ref{sec:analysis}, we describe the statistical analysis method for JUNO used in this work. In Section~\ref{sec:results}, we present the results of constraining and disentangling damping signatures at JUNO. We conclude in Section~\ref{sec:conclusions}.

\section{Damping signatures from new physics models}
\label{sec:models}
Damping signatures can be induced by a class of new physics models. 
Here, we focus on the exponential damping framework~\cite{Blennow:2005yk,Blennow:2006hd}, i.e., they can be written in the form of multiplying each term of the neutrino oscillation probabilities with exponential factors, which can arise from an approximation of the  first- or second-order perturbations to the standard neutrino oscillation probabilities from new physics scenarios~\cite{Liu:1997zd, Cheng:2020jje, Shafaq:2021lju}. 
	In this framework, the general expression for the probability of $\nu_a$ oscillating into $\nu_b$ in vacuum is given by
\begin{equation}
P(\nu_a \to \nu_b)= \sum_\text{i,j=1}^3U_{a\rm j}U^*_{b\rm j}U^*_{a\rm i}U_{b\rm i}\exp\left(-\mathbf{i}\frac{\Delta m^2_\text{ij}L}{2E}\right)D_\text{ij}(\alpha_{\rm ij}),
\label{eq:general_eq}\,
\end{equation}
where $U$ is the Pontecorvo-Maki-Nakagawa-Sakata (PMNS) mixing matrix~\cite{Giganti:2017fhf,Zyla:2020zbs}, $\Delta m^2_\text{ij}=m_\text{i}^2-m^2_\text{j}$, with $m_\text{i}$ being the eigenstate mass of $\nu_\text{i}$; $L$ is the baseline length, $E$ is the neutrino energy; $D_\text{ij}$ is an exponential damping factor and the specific form can be found in Table~\ref{tab:dampingmodels}, and the $\alpha_{\rm ij}$ are damping coefficients. Hereinafter, except for the $\nu_3$ decay case, we assume universal couplings, i.e., $\alpha_\text{ij}\equiv\alpha$, to describe the magnitudes of different damping effects.

\newcommand\rownumber{\stepcounter{magicrownumbers}(\arabic{magicrownumbers})}
\begin{table}
	\renewcommand\arraystretch{1.2}
	\begin{center}
		\resizebox{\textwidth}{!}
		%\caption{Different examples of damping signatures (suitable units)}
		{\begin{tabular}{|c|cccc|}
				\hline
				Type&Damping effect&Reference&Damping factor $D_\text{ij}$& Units of $\alpha$\\
				\hline
				\rownumber&\tabincell{c}{QD I}&\cite{Fogli:2007tx,Ribeiro:2007jq,Farzan:2008zv,Machado:2011tn,deOliveira:2013dia,Coelho:2017zes,Coloma:2018idr,Gomes:2020muc}& $\exp(-\alpha L/E^2)$ & $\mathrm{MeV^{2}\cdot m^{-1}}$\\
				\rownumber&\tabincell{c}{QD II}&\cite{Liu:1997km,Lisi:2000zt, Morgan:2004vv,Mavromatos:2006yy,Fogli:2007tx,Ribeiro:2007jq,Farzan:2008zv,Mehta:2011qb,Machado:2011tn,deOliveira:2013dia,Bakhti:2015dca,Gomes:2016ixi,Coelho:2017zes,Coloma:2018idr,Gomes:2020muc}& $\exp(-\alpha L)$ & $\mathrm{m^{-1}}$\\
				\rownumber &\tabincell{c}{QD III}&\cite{Lisi:2000zt, Morgan:2004vv, Blennow:2005yk,Mavromatos:2006yy,Fogli:2007tx,Ribeiro:2007jq,Farzan:2008zv,Mehta:2011qb,Machado:2011tn,deOliveira:2013dia,Bakhti:2015dca,Coelho:2017zes,Coloma:2018idr,Gomes:2020muc}& $\exp(-\alpha LE^2)$  & $\mathrm{MeV^{-2}\cdot m^{-1}}$\\
				\rownumber &Absorption&\cite{Blennow:2005yk,Fogli:2007tx,Ribeiro:2007jq,Farzan:2008zv,Machado:2011tn,deOliveira:2013dia,Gomes:2016ixi,Coelho:2017zes,Coloma:2018idr} & $\exp(-\alpha LE)$ & $\mathrm{MeV^{-1}\cdot m^{-1}}$\\
				\rownumber&$\nu_3$ decay&\cite{Gonzalez-Garcia:2008mgl,MINOS:2010fgd,Gomes:2014yua,Abrahao:2015rba,Pagliaroli:2016zab,Choubey:2018cfz,Ghoshal:2020hyo}&$\left\{\exp\left(-\alpha \frac{L}{E}\right),~\exp\left(-\alpha \frac{L}{2E}\right)\right\}$ & $\mathrm{MeV\cdot m^{-1}}$\\
				
				%\rownumber&Oscillation to $\nu_s$&\cite{Blennow:2005yk} & $\exp(-\alpha\frac{L^2}{(2E)^2})$ & $\mathrm{eV^4}$&2&2&0\\
				
				\rownumber&\tabincell{c}{WPD I}&\cite{Giunti:1991sx,Giunti:1997wq,Giunti:2003ax,Blennow:2005yk,Blasone:2015lya,Kersten:2015kio,Coelho:2017zes,deGouvea:2020hfl, deGouvea:2021uvg}& $\exp\left(-\alpha \frac{(\Delta m^2_\text{ij})^2L^2}{E^4}\right)$ & $\mathrm{MeV^2}$\\
				\rownumber&\tabincell{c}{WPD II}&\cite{Ohlsson:2000mj, Blennow:2005yk,Mavromatos:2006yy,Cheng:2020jje}& $\exp\left(-\alpha \frac{(\Delta m^2_\text{ij})^2L^2}{E^2}\right)$ & $\mathrm{dimensionless}$\\
				
				\rownumber&\tabincell{c}{WPD III}&\cite{Naumov:2010um,Chan:2015mca,An:2016pvi,Naumov:2020yyv,Cheng:2020jje}& $\exp(-R-\mathbf{i}X)$ & $\mathrm{dimensionless}$\\	
				
				%\rownumber&\tabincell{c}{QD IV}&\cite{Adler:2000vfa, Blennow:2005yk}& $\exp(-\alpha \frac{(\Delta m^2_\text{ij})^2L}{E^2})$ & $\mathrm{MeV^{-1}}$&1&2&2\\
				
				%\rownumber&\cite{Ohlsson:2000mj}& $\exp(-\alpha(\Delta m^2_{ij})^2)$ & $eV^{-4}$&0&0&2\\
				%\rownumber &\cite{Abrahao:2015rba, Porto-Silva:2020gma}& $\nu_3$ decay& $\exp(-\alpha \frac{L}{E}),\exp(-\alpha \frac{L}{2E})$ & $MeV\cdot m^{-1}$\\
				\hline
		\end{tabular}}
	\end{center}
	\caption{List of new physics models with different exponential damping factors. 
		%In general, the damping parameter $\alpha$ is a nonnegative damping coefficient matrix. Here it is assumed to be a nonnegative value for convenience in estimating the magnitude of the damping effect.
		The definitions of the parameters in the type (8) model are given in Eq.~(\ref{eq:WPD III_exp_form}).%$\alpha_{\epsilon}=4\alpha$ in the index (6), $\alpha_{\epsilon}=8\alpha$ in the index (8)
		% please see the main text. %Here, $\alpha=\sigma^2_E=1/(2\sigma_x)^2$ in the case (1), where $\sigma_x$ is the width of the neutrino wave packet. $\alpha=m_3/\tau_3$ in the case (2), where the 3rd mass eigenstate neutrino decays with lifetime $\tau_3$ at rest. $\alpha$ in the case (3) represents the magnitude of mixing a light sterile neutrino and three active neutrinos. $\alpha=\rho\sigma(E_0)/E_0$ in the case (4), where $\rho$ is the matter density, $\sigma(E_0)$ is the absorption cross-section at neutrino energy $E_0$. $\alpha$ in case (5-7), and $\eta_{\rm ij}=\frac{1}{2}\left(\frac{\Delta m^2_{\rm ij}}{4\sigma_{\rm rel}E^2}\right)^2$
		%; see Section~\ref{sec:signature} for more details.
		%$R$ and $X$ in the type (8) model are functions with parameters $\alpha,~\Delta m^2_{\rm ij},~L,~\mathrm{and}~E$ as independent variables, respectively.
		\label{tab:dampingmodels}}
\end{table}
\setcounter{magicrownumbers}{0}

The damping signatures from various new physics models are summarized in Table~\ref{tab:dampingmodels}. These models include quantum decoherence (QD), neutrino absorption, $\nu_3$ decay, and wave packet decoherence (WPD).
The new physics models of types (1) - (5) in Table~\ref{tab:dampingmodels} are expressed as power-law dependencies of the exponential form, i.e., $\exp(-\alpha LE^n)$ with $n=0$, $\pm1$, and $\pm2$~\cite{Lisi:2000zt,Mavromatos:2006yy, Fogli:2007tx, Ribeiro:2007jq, Farzan:2008zv,Machado:2011tn, deOliveira:2013dia, Bakhti:2015dca, Gomes:2016ixi, Coelho:2017zes, Coloma:2018idr, Gomes:2020muc}.  
Specifically, the type (1) model ($n=-2$) is demonstrated in Ref.~\cite{Fogli:2007tx} that it has the same functional form as the effects induced by stochastic density fluctuations. Thus, it is used to probe QD effects that might be induced by matter density fluctuations. The corresponding constraints of this model can be interpreted as limits on possible matter density fluctuations in the Sun~\cite{Fogli:2007tx}. The most significant feature of the type (2) model ($n=0$) is independent of neutrino energy. Many researchers have focused on this model since it is the simplest case of QD effects that might be induced by quantum gravity~\cite{Liu:1997km,Lisi:2000zt, Morgan:2004vv,Mavromatos:2006yy,Fogli:2007tx,Ribeiro:2007jq,Farzan:2008zv,Mehta:2011qb,Machado:2011tn,deOliveira:2013dia,Bakhti:2015dca,Gomes:2016ixi,Coelho:2017zes,Coloma:2018idr,Gomes:2020muc}. The type (3) model  ($n=2$)  is used to probe QD effects that might be induced by the space-time ``foam''  configurations of quantum gravity or D-brane of the form $\alpha\propto E^2/M_\text{Planck}$~\cite{Ellis:1996bz,Ellis:1997jw,Lisi:2000zt,Fogli:2007tx}, where $M_\text{Planck}$ is the Planck mass scale. The type (4) model ($n=1$), which is called neutrino absorption in Ref.~\cite{Blennow:2005yk}, is used to describe the absorption effect when neutrinos propagate through matter. In this type of model, $\alpha\equiv\rho\sigma(E_0)/E_0$, where $\rho$ is the matter density and $\sigma(E_0)$ is the effective cross section for neutrinos with an energy of $E_0$. Currently, neither atmospheric, solar neutrino oscillation experiments nor the long-baseline reactor neutrino experiment Kamioka Liquid Scintillator Anti-Neutrino Detector (KamLAND) shows evidence in favor of the new physics effects described by the previous four models ($n=0$, 1 and $\pm2$)~\cite{Lisi:2000zt,Fogli:2007tx}, which also indicates that their damping parameter $\alpha$ can be strongly constrained.  
%These constraints on $\alpha$ become stronger as the neutrino energies increase. 
Furthermore, there are no significant changes in the best-fit neutrino oscillation parameters in these new physics scenarios~\cite{Machado:2011tn, deOliveira:2013dia, Gomes:2016ixi, Gomes:2020muc}. The fact that neutrinos are massive implies they could decay. The $n=-1$ case was used in Refs.~\cite{Lindner:2001fx,Joshipura:2002fb,Beacom:2002cb,Fogli:2003th,Blennow:2005yk, Mehta:2011qb,Baerwald:2012kc,Abrahao:2015rba,Picoreti:2015ika,Gago:2017zzy,SNO:2018pvg,Porto-Silva:2020gma} to describe invisible neutrino decay scenarios, which lead to the violation of three-flavor neutrino unitarity. However, Ref.~\cite{Porto-Silva:2020gma} has shown that astrophysical neutrinos are potentially the most powerful source for constraining the decay parameters of $\nu_1$ and $\nu_2$, which could lead to the lower bounds on $\tau/m \sim 10^{-4}~(10^{6})$ $\mathrm{s/eV}$ from the solar (supernova) neutrinos. 
Nevertheless, the constraints on $\nu_3$ decay is much weaker than those on $\nu_1$ and $\nu_2$ from the current data~\cite{Abrahao:2015rba,Porto-Silva:2020gma,Ghoshal:2020hyo}. Here, in the type (5) model, we only consider  the $\nu_3$ decay scenario~\cite{Gonzalez-Garcia:2008mgl,MINOS:2010fgd,Gomes:2014yua,Abrahao:2015rba,Pagliaroli:2016zab,Choubey:2018cfz,Ghoshal:2020hyo}. The oscillation probability of $P(\bar{\nu}_e\rightarrow \bar{\nu}_e)$ comprises two exponential forms derived from the case of $n=-1$, with $\alpha$ being the neutrino eigenstate mass divided by the corresponding lifetime, i.e., $\alpha\equiv m_3/\tau_3$.

Although the plane-wave approximation theory successfully interprets a wide range of neutrino experiments, it is not self-consistent and leads to many paradoxes~\cite{Giunti:2003ax,Akhmedov:2009rb,Naumov:2010um,Naumov:2020yyv}. Therefore, the models of types (6) - (8) are proposed to form a consistent description of neutrino oscillations,  which use the wave packet treatment of neutrino oscillation instead of the plane wave approximation for neutrino propagation~\cite{Giunti:2003ax,Naumov:2010um,Chan:2015mca,An:2016pvi,Naumov:2020yyv,Cheng:2020jje}. However, this description also induce some WPD effects, which have not been found in current experimental data~\cite{An:2016pvi,deGouvea:2020hfl, deGouvea:2021uvg}.  Furthermore,  the WPD effects and $\nu_3$ decay can shift the best-fit neutrino oscillation parameters if these effects are strong enough~\cite{Blennow:2005yk,Abrahao:2015rba,An:2016pvi,Choubey:2018cfz,deGouvea:2020hfl,deGouvea:2021uvg}. Specifically, the type (6) model is used to describe the decoherence effect caused by wave packet separation~\cite{Giunti:1991sx,Giunti:1997wq,Giunti:2003ax,Blennow:2005yk,Kersten:2015kio,Blasone:2015lya,Coelho:2017zes,deGouvea:2020hfl, deGouvea:2021uvg}. This effect is related to the characteristics of the neutrino source and detector. In the type (6) model, $\alpha\equiv1/(4\sqrt{2}\sigma_x)^2$, where $\sigma_x$ is the spatial width of the neutrino wave packet. The type (7) model is used in Ref.~\cite{Ohlsson:2000mj} to show that in the two-neutrino oscillation case, a Gaussian-averaged neutrino oscillation model with $\exp[-2\sigma^2(\Delta m^2)^2]$ and a neutrino decoherence model with $\exp(-d^2L)$ are equivalent if $d=\frac{\sqrt{2}\Delta m^2}{\sqrt{L}}\sigma$ is fulfilled, where $\sigma$ is the standard deviation of $L/E$ and $d$ is the decoherence parameter. The model with $\exp[-2\sigma^2(\Delta m^2)^2]$ is obtained by Gaussian average over the $L/E$ dependence for the oscillation probability under the plane-wave approximation due to uncertainties in the energy and oscillation length~\cite{Ohlsson:2000mj,Mavromatos:2006yy}.
Since under the condition of $(2\sigma^2 E^4/L^2)=1/(4\sqrt{2}\sigma_x)^{2}$, the type (6) and type (7) models are equivalent, we refer to the type (7) model as WPD II. 

The type (8) model systematically studies the quantum decoherence effects caused by wave packet separation, dispersion and delocalization. We rewrite the unified decoherence effect in exponential form to discuss its impact on the neutrino oscillation probability. This exponential damping factor is given by~\cite{Chan:2015mca,Naumov:2010um,An:2016pvi,Naumov:2020yyv,Cheng:2020jje}
\beq
\begin{split}
\exp(-R-\mathbf{i}X)&=\exp\left\{-\left[\frac{1}{4}\ln(1+y^2_{\rm ij})+\lambda_{\rm ij}+\eta_{\rm ij}\right]-\mathbf{i}\left[\frac{1}{2}\tan^{-1}(y_{\rm ij})-\lambda_{\rm ij}y_{\rm ij}\right]\right\}\\&=\left(\frac{1}{1+y^2_{\rm ij}}\right)^{\frac{1}{4}}\exp(-\lambda_{\rm ij})\exp\left(-\frac{\mathbf{i}}{2}\tan^{-1}(y_{\rm ij})\right)\exp(\mathbf{i}\lambda_{\rm ij}y_{\rm ij})\exp(-\eta_{\rm ij})
\label{eq:WPD III_exp_form},
\end{split}
\eeq
where $\lambda_{\rm ij}=\frac{x^2_{\rm ij}}{1+y^2_{\rm ij}}$, $x_{\rm ij}=\frac{\sqrt{2}\Delta m^2_{\rm ij}L}{4E}\sigma_{\rm rel}$, $y_{\rm ij}=\frac{\Delta m^2_{\rm ij}L}{E}\sigma^2_{\rm rel}$, $\eta_{\rm ij}=\frac{1}{2}\left(\frac{\Delta m^2_{\rm ij}}{4\sigma_{\rm rel}E^2}\right)^2$, and $\sigma_{\rm rel}=(2\sigma_xE)^{-1}$.
In this model, we define $\alpha\equiv\sigma_{\rm rel}$, where $\sigma_\text{rel}$ represents the intrinsic relative dispersion of neutrino momentum. The $\exp(-\lambda_{\rm ij})$  therm corresponds to the conventional quantum decoherence effect caused by the gradual separation of different mass states traveling at different spatial propagation speeds, which causes them to stop interfering with each other, leading to damped oscillations. The terms containing $y_{\rm ij}$ describe the dispersion effect, which includes two effects on the oscillations: wave packet spreading compensates for wave packet separation, and dispersion reduces the overlap fraction of the wave packets~\cite{Chan:2015mca,Cheng:2020jje}. %Eq.~(\ref{eq:WPD III_exp_form}) can be multiplied by an $\exp(-\eta_{\rm ij})$ term to include the decoherence effect caused by delocalization~\cite{Chan:2015mca,An:2016pvi,Cheng:2020jje,Naumov:2020yyv,Naumov:2010um}, where $\eta_{\rm ij}=\frac{1}{2}\left(\frac{\Delta m^2_{\rm ij}}{4\sigma_{\rm rel}E^2}\right)^2$. 
The $\exp(-\eta_{\rm ij})$ term corresponds to the quantum decoherence effect from delocalization, which is related to the neutrino production and detection processes and is independent of the baseline $L$. We find that $\exp(-\eta_{\rm ij})$ is very close to 1 at JUNO if $\sigma_{\rm rel}\gtrsim\rm O(10^{-15})$. In Ref.~\cite{An:2016pvi}, the Daya Bay (DYB) collaboration published their first experimental limits, which are $10^{-14}<\sigma_{\rm rel}<0.23$ and $2.38\times10^{-17}<\sigma_{\rm rel}<0.23$ at a 95\% CL when the dimensions of the reactor cores and detectors are and are not considered as constraints, respectively. Therefore, we neglect the $\exp(-\eta_{\rm ij})$ term in Eq.~(\ref{eq:WPD III_exp_form}) in this work in the following text\footnote{If we consider the decoherence effect caused by delocalization, the lower limit on $\sigma_{\rm rel}$ at JUNO can reach $3.0\times10^{-17}$ at 95\% CL. Although this expected lower limit is slightly better than the DYB limit of $\sigma_{\rm rel}>2.38\times10^{-17}$, the improvement from JUNO is not large due to the smaller IBD events compared with DYB and the baseline independence of delocalization~\cite{An:2016pvi}. }.

In addition, some works have discussed exponential damping models such as $\exp\left(-\alpha\frac{L^2}{(2E)^2}\right)$ and $\exp\left(-\alpha \frac{(\Delta m^2_\text{ij})^2L}{E^2}\right)$. The former was adopted in Ref.~\cite{Blennow:2005yk} to approximately describe the mixing of three active neutrinos and a very light sterile neutrino in short-baseline reactor neutrino experiments. Here, $\alpha$ represents the magnitude of mixing between the three active neutrinos and the light sterile neutrino. Note that this approximate relationship does not hold for medium- or long-baseline neutrino experiments with an eV-scale sterile neutrino or for mixing scenarios involving three active neutrinos and multiple sterile neutrinos. The latter damping model was proposed to explain the decoherence effect caused by quantum gravity in the Super-Kamiokande experiment~\cite{Adler:2000vfa}, and the coupling $\alpha$ can be related to $M_\text{Planck}$. %the Planck mass scale. 
For a single-baseline experiment or an experiment with multiple identical baselines, the phenomenology of the former model above is the same as that of the type (1) model, and the phenomenology of the latter model above is the same as that of the type (7) model. Therefore, we will not discuss these two models in depth in this paper.

%In summary, we might as well classify the case   (1), (2), and (3) model into one category in the following study of this paper.
%such as JUNO,  with exponential damping factors

%%%%%%%%%%%%%%%%%%%%%%%%%%%%%
\section{Damping signatures at medium-baseline reactor neutrino experiments}
\label{sec:dampingeffects}
In this section, we first discuss the damping effects on the survival probability of $\bar{\nu}_e$ in medium-baseline reactor neutrino experiments. After that, we classify the damping effects in accordance with their different damping behaviors.%distortion performance on the survival probability spectrum of $\bar{\nu}_e$.
\subsection{Damped neutrino oscillation probabilities} 
\label{subsec:framework}
From the general expression in Eq.~(\ref{eq:general_eq}), we can obtain four cases for the damped survival probability of reactor neutrinos ($\bar{\nu}_e$) in vacuum, as follows:

\begin{enumerate}[label=(\Roman*)]
\item The overall $\bar{\nu}_e$ survival probability is damped out. This case includes the QD I, QD II, QD III, and absorption damping effects. %In Eq.~(\ref{eq:qd1damping-reactor}), the expression in curly brackets represents the survival probability of reactor neutrinos in a vacuum without damping effects, and $D=D_{\rm ij}$ because of without $\Delta m^2_{\rm ij}$.
\beq
\begin{split}
P(\bar{\nu}_e \to \bar{\nu}_e)= D\{1&-c^4_{13}\sin^2(2\theta_{12})\sin^2(\Delta_{21})-c^2_{12}\sin^2(2\theta_{13})\sin^2(\Delta_{31})\\&-s^2_{12}\sin^2(2\theta_{13})\sin^2(\Delta_{32})\},
	\label{eq:qd1damping-reactor}
\end{split}
\eeq
where the expression in curly brackets represents the $\bar{\nu}_e$ survival probability in vacuum without damping effects (i.e., the standard $\bar{\nu}_e$ survival probability), $D=D_{\rm ij}$ because there are no relevant $\Delta m^2_{\rm ij}$ terms in these damping factors, $c_\text{ij}=\cos\theta_\text{ij}$, $s_\text{ij}=\sin\theta_\text{ij}$, and the oscillation phase $\Delta_\text{ij}$ is defined as
\beq
\begin{split}
\Delta_\text{ij}=\frac{\Delta m^2_\text{ij}L}{4E}\simeq 1.267\frac{\Delta m^2_\text{ij}[\mathrm{eV^2}]L[\mathrm{km}]}{E[\mathrm{GeV}]}=1.267\frac{\Delta m^2_\text{ij}[\mathrm{eV^2}]L[\mathrm{m}]}{E[\mathrm{MeV}]}.
	\label{eq:delta_ij}\,
\end{split}
\eeq

\item  Some oscillating and nonoscillating terms of the $\bar{\nu}_e$ survival probability are damped out. This case includes the $\nu_3$ decay damping effect.
\beq
\begin{split}
P(\bar{\nu}_e \to \bar{\nu}_e)= &c^4_{13}[1-\sin^2(2\theta_{12})\sin^2(\Delta_{21})]\\
&+\frac{1}{2}\sin^2(2\theta_{13})\exp(-\frac{\alpha L}{2E})[c^2_{12}\cos(2\Delta_{31})+s^2_{12}\cos(2\Delta_{32})]+\exp(-\frac{\alpha L}{E})s^4_{13}.
\label{eq:v3decaydamping-reactor}
\end{split}
\eeq

\item Only the oscillating terms of the $\bar{\nu}_e$ survival probability are damped out, but there are no dispersion terms. This case includes the WPD I and WPD II damping effects.
\beq
\begin{split}
%P(\bar{\nu}_e \to \bar{\nu}_e)&= c^4_{13}\{1-\frac{1}{2}\sin^2(2\theta_{12})[1-D_{21}\cos(2\Delta _{21})]\}\\
%&+\frac{1}{2}\sin^2(2\theta_{13})[D_{31}c^2_{12}\cos(2\Delta_{31})+D_{32}s^2_{12}\cos(2\Delta_{32})]+s^4_{13}
P(\bar{\nu}_e \to \bar{\nu}_e)= 1&-\frac{1}{2}[c^4_{13}\sin^2(2\theta_{12})+\sin^2(2\theta_{13})]+\frac{1}{2}c^4_{13}\sin^2(2\theta_{12})D_{21}\cos(2\Delta _{21})\\
&+\frac{1}{2}\sin^2(2\theta_{13})[D_{31}c^2_{12}\cos(2\Delta_{31})+D_{32}s^2_{12}\cos(2\Delta_{32})].
	\label{eq:wpd1damping-reactor}
\end{split}
\eeq

\item Not only are the oscillating terms of the $\bar{\nu}_e$ survival probability damped out, but there are also dispersion terms. This case includes the WPD III damping effect.

\beq
\begin{split}
%\begin{equation}
%P(\nu_a \to \nu_b)\approx \sum_\text{i,j=1}^3\left\{U_{a\rm j}U^*_{b\rm j}U^*_{a\rm i}U_{b\rm i}\exp\left(-\mathbf{i}\frac{\Delta m^2_\text{ij}L}{2E}\right)\right\}\left\{\left(\frac{1}{1+y_{\rm ij}}\right)^{\frac{1}{4}}\exp(-\lambda_{\rm ij})\exp\left(\frac{-\mathbf{i}}{2}\tan^{-1}(y_{\rm ij})\right)\exp(\mathbf{i}\lambda_{\rm ij}y_{\rm ij})\right\}
P(\bar{\nu}_e \to \bar{\nu}_e)=1&-\frac{1}{2}c_{13}^4\sin^2(2\theta_{12})\left[1-\left(\frac{1}{1+y^2_{21}}\right)^{\frac{1}{4}}\exp(-\lambda_{21})\cos(\phi_{21})\right]\\
&-\frac{1}{2}\sin^2(2\theta_{13})c^2_{12}\left[1-\left(\frac{1}{1+y^2_{31}}\right)^{\frac{1}{4}}\exp(-\lambda_{31})\cos(\phi_{31})\right]\\
&-\frac{1}{2}\sin^2(2\theta_{13})s^2_{12}\left[1-\left(\frac{1}{1+y^2_{32}}\right)^{\frac{1}{4}}\exp(-\lambda_{32})\cos(\phi_{32})\right],
\label{eq:whole_wp}
\end{split}
\eeq
where $\phi_{\rm ij}=\frac{\Delta m^2_{\rm ij}L}{2E}+\frac{1}{2}\arctan(y_{\rm ij})-\lambda_{\rm ij}y_{\rm ij}$ and is the sum of the plane wave phase and the phase shift introduced by wave packet dispersion.
\end{enumerate}

In general, the $\bar{\nu}_e$ survival probability at JUNO is also affected by the Mikheyev--Smirnov--Wolfenstein (MSW) matter effect as the neutrinos travel through matter~\cite{Wolfenstein:1977ue, Mikheyev:1985zog}. We can treat this damping effect as a minor perturbation of the neutrino oscillations in matter~\cite{Blennow:2005yk}. For the standard three-neutrino oscillation scenarios, the corrections to the neutrino parameters due to matter effects do not exceed 1.1\%~\cite{An:2015jdp, Li:2016txk,Khan:2019doq}. In this work, we also ignore matter effects because they only slightly shift the central values of the neutrino oscillation parameters and do not affect the measurement precision.

\subsection{Classification of damping effects}
\label{subsec:classification}

\begin{figure}[htb]
	\centering
	\includegraphics[width=\textwidth]{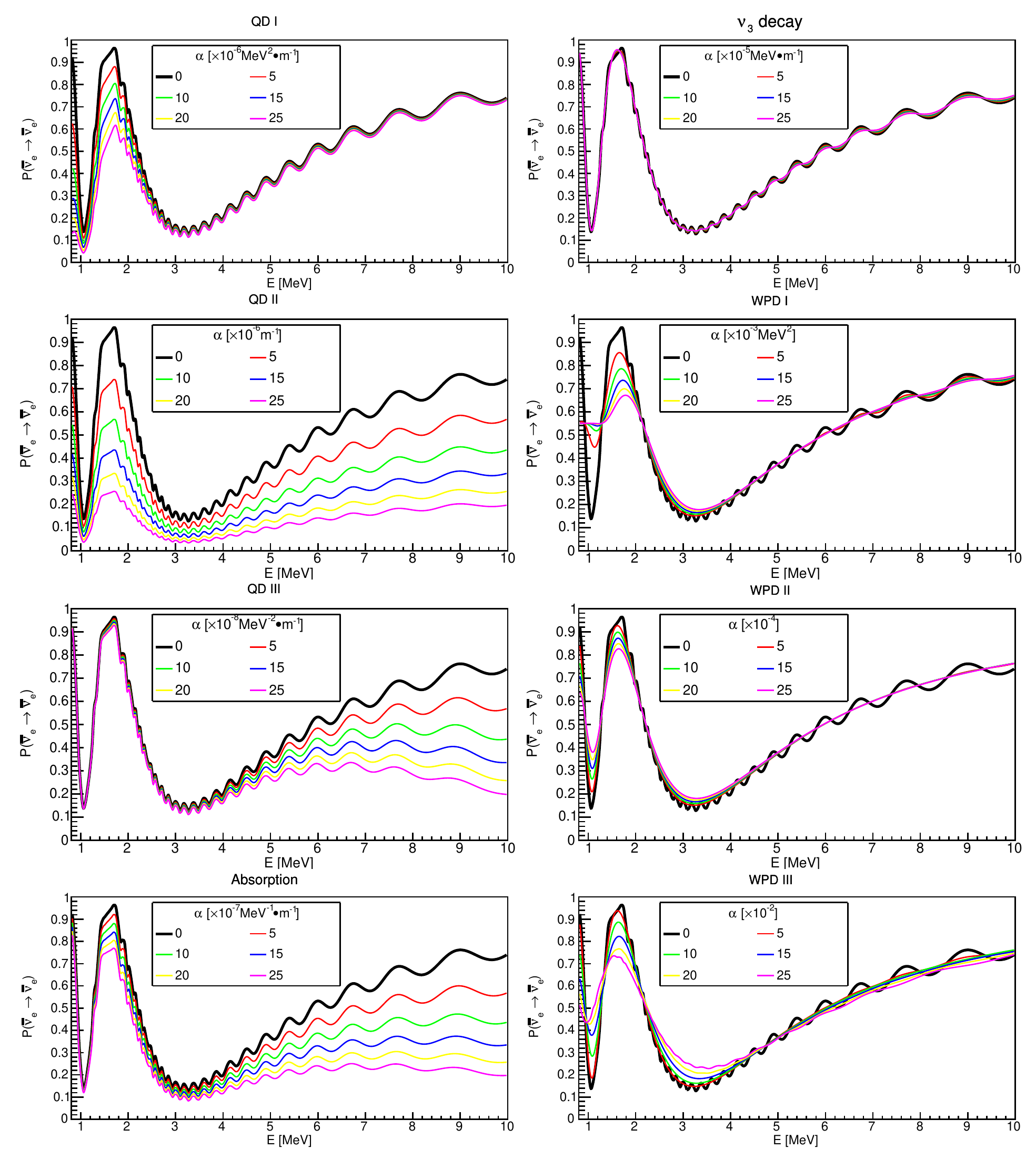}
\caption{ The $\bar{\nu}_e$ survival probability $P(\bar{\nu}_e\rightarrow \bar{\nu}_e)$ with different damping parameter values for each new physics model. 
	\label{fig:possibilities}}
\end{figure}

In Figure~\ref{fig:possibilities}, we plot the $\bar{\nu}_e$ survival probability $P(\bar{\nu}_e\rightarrow \bar{\nu}_e)$ with different damping parameter values for each new physics model. The neutrino oscillation parameters are taken from Ref.~\cite{Zyla:2020zbs} and summarized in Table \ref{tab:pdg2020}. We assume the NMO in this analysis. We find that the results are quite similar for the IMO. We choose a few values for the damping parameters for illustration. In particular, $\alpha=0$ indicates no damping effect, i.e., neutrino oscillation of the standard type. The farther the spectrum is from the no-damping curve, the stronger the intensity of the damping effect.
The distortion of the standard $\bar{\nu}_e$ survival probability spectrum caused by damping is a combined phenomenon of an amplitude decrease and a phase shift, which can be regarded as a unique signature, as shown in Figure ~\ref{fig:possibilities}. We find that the amplitude decrease behaviors of both the fast oscillation cycles (driven by $\Delta m^2_{31}$ and $\Delta m^2_{32}$) and the slow oscillation cycles (driven by $\Delta m^2_{21}$) are more significant than their phase shift behaviors in all damping effect scenarios. Therefore, damping effects mainly smear the fine structure of the standard $\bar{\nu}_e$ survival probability spectrum through amplitude-decreasing effects. Furthermore, the fine structure of the fast oscillation cycles is smeared more strongly than that of the slow oscillation cycles with increasing $\alpha$, which indicates that more spectral shape information is lost in the former than in the latter.

Based on the different smearing behaviors, we can divide the damping effects in Table~\ref{tab:dampingmodels} into three categories. The first category is referred to as the QD-like effects, which include the QD I, QD II, QD III, and absorption damping effects. Although the details of the smearing behavior of each model are different, the fine structure is more completely preserved under increasing $\alpha$ for models in this category than for models in the other two categories. As $\alpha \to \infty$, the $\bar{\nu}_e$ survival probabilities of the models in this category approach zero, which means that the neutrinos do not propagate. The second category includes the $\nu_3$ decay effect. In this category, the fine structure of the fast oscillation cycles will be smeared more strongly as $\alpha$ increases until all details of the fast oscillation structure are lost. However, the damping effects of this category will not affect the fine structure of the slow oscillation cycles. Consequently, only the slow oscillation cycles will remain as $\alpha \to \infty$. The third category is referred to as WPD-like effects, which include the WPD I, WPD II, and WPD III damping effects. As $\alpha$ increases, the fine structures of both the fast and slow oscillation cycles will be strongly smeared under WPD-like effects, but the former will be smeared out before the latter. The $\bar{\nu}_e$ survival probabilities of these models approach a nonzero constant value as $\alpha \to \infty$, i.e., $1-\frac{1}{2}[c^4_{13}\sin^2(2\theta_{12})+\sin^2(2\theta_{13})]$. Notably, the number of neutrinos will be lost in the damping models of the first and second categories, whereas they will keep the same in the third category.

%We can divide the damping effects in Table~\ref{tab:dampingmodels} into two categories depending on whether neutrinos are lost in flavor transitions. The category of neutrino loss includes the QD I, QD II, QD III, absorption, and $\nu_3$ decay damping effects. The other category, which corresponds to scenarios without neutrino loss, includes the WPD I, WPD II, and WPD III damping effects.

\begin{table}
\begin{center}\renewcommand\arraystretch{1.5}
%\caption{PDG 2020}
{\begin{tabular}{cccccc}
\hline
$p$& $\sin^2\theta_{12}$&$\sin^2\theta_{13}$&$\Delta m^2_{21}(\mathrm{eV^2})$&$\Delta m^2_{32}(\mathrm{NMO,eV^2})$&$\Delta m^2_{32}(\mathrm{IMO,eV^2})$\\
\hline
$p^\text{input}$&0.307 & $2.18\times10^{-2}$ & $7.53\times10^{-5}$ & $2.453\times10^{-3}$& $-2.546\times10^{-3}$\\
$\delta p$&0.013&$0.07\times10^{-2}$&$0.18\times10^{-5}$&$0.034\times10^{-3}$&$0.037\times10^{-3}$\\
\hline
\end{tabular}}
\end{center}
\caption{ The neutrino oscillation parameters used in this work~\cite{Zyla:2020zbs}. % Here, NMO denotes the normal mass ordering, and IMO denotes the inverted mass ordering.
 The input values $p^\text{input}$ and the corresponding $1\sigma$ uncertainty values $\delta p$ are taken from Ref.~\cite{Zyla:2020zbs}. For the case in which $\Delta m^2_{32}$ is negative, the corresponding $\delta p$ is the average value.
\label{tab:pdg2020}}
\end{table}

%according to the form of Eq. (\ref{eq:newdecaydamping-reactor}) and
%%%%%%%%%%%%%%%%%%%%%%%%%%%%%%%%
%\section{Constraints on the damping parameters at JUNO}
\section{Analysis method for JUNO}
\label{sec:analysis}
The damping effects on the reactor neutrino oscillations can be probed at JUNO by measuring the distortion of the neutrino inverse beta decay (IBD) event spectrum. The observed $\bar{\nu}_e$ distribution in terms of the reconstructed energy ($E_\text{rec}$) can be expressed as follows~\cite{Ge:2012wj}:
\beq
\begin{split}
\frac{dN}{dE_\text{rec}}=\frac{N_\text{p}T}{4\pi L^2}\int_{m_n-m_p+m_e}dE\frac{W_{\rm th}}{\sum_uf_u\varpi_u}\sum_uf_uS_u(E)  P(\bar{\nu}_e \to \bar{\nu}_e)\sigma_\text{IBD}(E)G(E_\text{vis}-E_\text{rec},\delta E_\text{vis}),
	\label{eq:event_rates}\,
\end{split}
\eeq
where $N_\text{p}$ is the total number of free target protons in the LS detector, $T$ is the total exposure time, and $W_{\rm th}$ is the thermal power of the reactor. $f_u$, $\varpi_u$, and $S_u$ are the fission fraction, the mean energy released per fission, and the $\bar{\nu}_e$ energy spectrum per fission, respectively, for the isotope $u$, where $u=\{\rm ^{235}U,~^{238}U,~^{239}Pu,~^{241}Pu\}$. The values of $f_u$ and $\varpi_u$ are taken from Ref.~\cite{DayaBay:2016ssb}. $S_{\rm ^{235}U}$, $S_{\rm ^{239}Pu}$, and $S_{\rm^{241}Pu}$ are derived from Ref.~\cite{Huber:2011wv}, and $S_{\rm ^{238}U}$ is derived from Ref.~\cite{Mueller:2011nm}. $\sigma_\text{IBD}(E)$ is the cross section for IBD in a detector, taken from Refs.~\cite{Strumia:2003zx,Dighe:2003be}; $E_\text{vis}$ is the visible energy ($E_\text{vis} \sim E_e+m_e \sim (E-0.8)~\mathrm{MeV}$), and $G(E_\text{vis}-E_\text{rec},\delta E_\text{vis})$ is a normalized Gaussian function representing a detector response function with an energy resolution of $\delta E_\text{vis}$. This function is expressed as follows:
\beq
\begin{split}
G(E_\text{vis}-E_\text{rec},\delta E_\text{vis})\approx\frac{1}{\sqrt{2\pi}\delta E_\text{vis}}\exp\left\{-\frac{(E_\text{vis}-E_\text{rec})^2}{2(\delta E_\text{vis})^2}\right\},
	\label{eq:gauss}\,
\end{split}
\eeq
where $\delta E_\text{vis}$ is taken from Ref.~\cite{An:2015jdp}. The detector energy resolution can be described by a three-parameter function, i.e.,
\beq
\begin{split}
\frac{\delta E_\text{vis}}{E_\text{vis}}=\sqrt{\left(\frac{p_0}{\sqrt{E_\text{vis}/\mathrm{MeV}}}\right)^2+p_1^2+\left(\frac{p_2}{E_\text{vis}/\mathrm{MeV}}\right)^2},
	\label{eq:resolution}\,
\end{split}
\eeq
where the parameters $p_0$, $p_1$ and $p_2$ represent the contributions to the energy resolution from the photon statistics, detector-related residual energy nonuniformity, and photomultiplier tube (PMT)-related effects, respectively.

The effective energy resolution of 3\% at 1 MeV of the JUNO detector, as discussed in Refs.~\cite{Abusleme:2020lur,Abusleme:2021zrw}, is considered, and we set $p_0=2.61\%$, $p_1=0.82\%$, and $p_2=1.23\%$. We also take the IBD detection efficiency of the detector to be 73\%~\cite{An:2015jdp,Abusleme:2020lur}. The JUNO detector is located at equal distances of $\sim53~\mathrm{km}$ from the Yangjiang and Taishan thermal power reactor complexes~\cite{An:2015jdp,Abusleme:2020lur,Abusleme:2021zrw}. The thermal powers of these two reactor complexes are $17.4~\mathrm{GW_{th}}$ and $9.2~\mathrm{GW_{th}}$, respectively~\cite{Abusleme:2020lur}. We consider the exposure of the JUNO detector to be $(26.6\times20\times6\times300)\ \mathrm{GW_{th}\cdot kton \cdot years \cdot days}$ and assume the NMO scenario unless explicitly stated otherwise.

For the analysis, we adopt the least square method from Refs.~\cite{Ge:2012wj, Capozzi:2013psa, Abrahao:2015rba, An:2015jdp, Wang:2016vua,Porto-Silva:2020gma} and define a $\chi^2$ function with proper nuisance parameters and penalty terms to quantify the sensitivity of $\alpha$, as follows:

\beq %\large% \tiny,\scriptsize , \footnotesize,\small,\normalsize,\large,\Large
\begin{large}
\begin{aligned}
  %\chi^2&=\chi^2_\text{stat}+\chi^2_\text{sys}+\chi^2_\text{param}\\
 \chi^2=&\sum^{N_\text{bin}}_i \frac{[M_i-T_i(1+\epsilon_R+\epsilon_d+\sum_r\omega_r\epsilon_r+\epsilon_s)-\sum_b B_{b,i}(1+\epsilon_b)]^2}{T_i+(\sigma^\text{shape} T_i)^2+\sum_b(B_{b,i}\sigma^\text{shape}_{b})^2}\\ 
 &+\frac{\epsilon^2_R}{\sigma^2_R}+\frac{\epsilon^2_d}{\sigma^2_d}+\sum_r\frac{\epsilon^2_r}{\sigma^2_r}+\frac{\epsilon_s^2}{\sigma^2_s}+\sum_b\frac{\epsilon_b^2}{\sigma_b^2}\\
 &+\sum_k\left(\frac{p^\text{input}_k-p^\text{fit}_k}{\delta p_k}\right)^2,
 \label{eq:chi2-full}\,
 \end{aligned}
 \end{large}
\eeq
where $N_\text{bin}$ is the number of energy bins, $M_i$ is the number of measured total events (the summation of signal and background) in the $i$-th bin, $T_i$ is the predicted number of IBD events, $B_b$ is the $b$-th kind of estimated background (the main background spectra for the JUNO detector are taken from Ref.~\cite{An:2015jdp}), and the quantities $\sigma$ and $\epsilon$ with different indices represent systematic uncertainties and the corresponding pull parameters, respectively. The considered systematic uncertainties include the correlated reactor uncertainty ($\sigma_R$=2\%), the detector-related uncertainty ($\sigma_d$=1\%), the uncorrelated reactor uncertainty ($\sigma_r$=0.8\%), the uncorrelated spectrum shape uncertainty ($\sigma_s$=1\%), the correlated spectrum shape uncertainty ($\sigma^\text{shape}$=1\%), the shape uncertainties of the backgrounds ($\sigma^\text{shape}_b$), and the relative rate uncertainties of the backgrounds ($\sigma_b$). Specifically, the $\sigma^\text{shape}_b$ values for accidental coincidences, fast neutrons, $\mathrm{^9Li/^8He}$, $\mathrm{^{13}C(\alpha,n)^{16}O}$ and geoneutrinos at JUNO are negligible (i.e., 0\%), 20\%, 10\%, 50\%, and 5\%, respectively; the corresponding $\sigma_b$ values are 1\%, 100\%, 20\%, 50\%, and 30\%, respectively. Additionally, $\omega_r$ is a fraction representing the $r$-th reactor's contribution to the corresponding pull parameter $\epsilon_r$. Finally, $p_k$ and $\delta p_k$ denote the $k$-th neutrino oscillation parameter ($\sin^2\theta_{12},~\sin^2\theta_{13},~\Delta m^2_{21},~\rm{or}~\Delta m^2_{32}$) and the corresponding uncertainty, respectively, at a $1\sigma$ CL; these values are given in Table~\ref{tab:pdg2020}.

\section{Results}
\label{sec:results}
In this section, we present the results of probing the damping signatures of different new physics models at JUNO.
We firstly study the constraints on the damping parameters for the eight new physics models at JUNO. Then, we show that JUNO can also help to disentangle the damping model from each other.

\subsection{Constraints on the damping parameters at JUNO}
To obtain the constraints on the damping parameters at JUNO, we scan the damping parameter of each damping model by marginalizing over other parameters, and fit the simulated no-damping JUNO data to obtain the exclusion sensitivities of the damping parameters. We list the constraints on the damping parameter of each damping model from this work in Table~\ref{tab:bounds}. The current bounds on the damping parameters in the literature are also listed for comparison.
The damping factors of the first seven damping models in Table~\ref{tab:bounds} can be unified into a general form~\cite{Blennow:2005yk,Blennow:2006hd},
\beq
\begin{split}
D_\text{ij}=\exp\left(-\alpha\frac{|\Delta m^2_\text{ij}|^\xi L^\beta}{E^\gamma}\right)
	\label{eq:damping-factors},
\end{split}
\eeq
where the parameters $\xi$, $\beta$, and $\gamma$ are the power numbers in the damping factor of interest. The strength of neutrino oscillation  experiments to probe the damping effects is strongly dependent on the specific values of $\xi$, $\beta$, and $\gamma$~\cite{Blennow:2005yk,Blennow:2006hd}. 

\begin{table}\renewcommand\arraystretch{1.5}%\scriptsize\scriptsize
\resizebox{\textwidth}{!}%KamLAND L= 180 km, E_classic=3.4 MeV  PhysRevD.88.033001  PhysRevLett.94.081801
{\begin{tabular}{|c|c|c|}
\hline
\tabincell{c}{Damping type\\Parameter $[\rm units]$}&\tabincell{c}{Phenomenological limits (experiment: original results, CL~[Ref])\\\{Experimental limits (experiment: original results, CL~[Ref])\}}&\tabincell{c}{Exclusion sensitivities \\for JUNO (CL)}\\
\hline
\tabincell{c}{QD I\\$\alpha$ $[\mathrm{\times10^{-6}~\frac{MeV^2}{m}}]$}&\tabincell{c}{$<1.62\times10^5$ (MINOS+T2K+reactor:~$\alpha<3.2\times10^{-23}~\mathrm{GeV^3}$, 90\%~\cite{Gomes:2020muc})\\$<0.41$ (solar+KL:~$\alpha<0.81\times10^{-28}~\mathrm{GeV^3}$, 95\%~\cite{Fogli:2007tx})}&\tabincell{c}{$<3.72$  (90\%)\\$<4.42$ (95\%)}\\% n=-2
\hline
\tabincell{c}{QD II\\$\alpha~[\times\mathrm{\frac{10^{-6}}{m}}]$}&\tabincell{c}{$<3.45$ (KL: $6.8\times10^{-22}~\mathrm{GeV}$, 95\%~\cite{Gomes:2016ixi})\\$<0.33$ (MINOS+T2K+reactor:~$\alpha<6.5\times10^{-23}~\mathrm{GeV}$, 90\%~\cite{Gomes:2020muc})\\$<0.18$ (SK:~$\alpha<3.5\times10^{-23}~\mathrm{GeV}$, 90\%~\cite{Lisi:2000zt})\\$<3.40\times10^{-3}$ (solar+KL:~$\alpha<0.67\times10^{-24}~\mathrm{GeV}$, 95\%~\cite{Fogli:2007tx})}&\tabincell{c}{$<0.80$ (90\%)\\$<0.95$ (95\%)}\\% n=0 
%%\\$<1.1\times10^{-23}$ (Kamioka+Korea: 90\%~\cite{Ribeiro:2007jq})
\hline
\tabincell{c}{QD III\\$\alpha~[\times\mathrm{\frac{10^{-8}}{MeV^{2}\cdot m}}]$}&\tabincell{c}{$<2.38\times10^{-3}$(solar+KL:~$\alpha<0.47\times10^{-20}~\mathrm{GeV^{-1}}$, 95\%~\cite{Fogli:2007tx})\\$<1.42\times10^{-5}$ (MINOS+T2K+reactor:~$\alpha<2.8\times10^{-23}~\mathrm{GeV^{-1}}$, 90\%~\cite{Gomes:2020muc})\\$<4.56\times10^{-10}$ (SK:~$\alpha<0.9\times10^{-27}~\mathrm{GeV^{-1}}$, 90\%~\cite{Lisi:2000zt})}&\tabincell{c}{$<1.22$ (90\%)\\$<1.46$ (95\%)}\\% n=2
%%\\$<1.7\times10^{-23}$ (Kamioka+Korea: 90\%~\cite{Ribeiro:2007jq})
\hline
\tabincell{c}{Absorption\\$\alpha~[\times\mathrm{\frac{10^{-7}}{MeV\cdot m}}]$}&\tabincell{c}{$<7.60$ (KL:~$\alpha<1.5\times10^{-19}$, 95\%~\cite{Gomes:2016ixi})\\$<0.10$ (SK:~$\alpha<2.0\times10^{-21}$, 90\%~\cite{Lisi:2000zt})\\$<2.94\times10^{-3}$ (solar+KL:~$\alpha<0.58\times10^{-22}$, 95\%~\cite{Fogli:2007tx})}&\tabincell{c}{$<1.04$ (90\%)\\$<1.23$ (95\%)}\\%n=1
%%\\$<1.7\times10^{-23}$ (Kamioka+Korea: 90\%~\cite{Ribeiro:2007jq})
\hline
\tabincell{c}{$\nu_3$ decay\\$\alpha\equiv\frac{m_3}{\tau_3}$\\$[\times10^{-4}~\mathrm{\frac{MeV}{m}}]$}&\tabincell{c}{$<256.59$ (OPERA:~$\frac{\tau_3}{m_3}>1.3\times10^{-13}~\mathrm{\frac{s}{eV}}$, 90\%~\cite{Pagliaroli:2016zab})\\$<22.24$ (NO$\nu$A+T2K:~$\frac{\tau_3}{m_3}>1.5\times10^{-12}~\mathrm{\frac{s}{eV}}$, 90\%~\cite{Choubey:2018cfz})\\$<0.36$ (SK+K2K+MINOS:~$\frac{\tau_3}{m_3}>9.3\times10^{-11}~\mathrm{\frac{s}{eV}}$, 99\%~\cite{Gonzalez-Garcia:2008mgl})\\\{$<15.88$ (MINOS:~$\frac{\tau_3}{m_3}>2.1\times10^{-12}~\mathrm{\frac{s}{eV}}$, 90\%~\cite{MINOS:2010fgd})\}}&\tabincell{c}{$<0.44$ (90\%)\\$<0.53$ (95\%)\\$<0.75$ (99\%)}\\
%%\\$>5.1\times10^{-11}$ (DUNE: 90\%~\cite{Ghoshal:2020hyo})
\hline
\tabincell{c}{WPD I\\$\alpha\equiv(4\sqrt{2}\sigma_x)^{-2}$\\$[\times\mathrm{10^{-3} MeV^{2}}]$}&\tabincell{c}{$<116.96$ (RENO+DYB:~$\sigma_x>1.02\times10^{-4}~\mathrm{nm}$, 90\%~\cite{deGouvea:2020hfl})\\$<27.59$ (RENO+DYB+KL:~$\sigma_x>2.1\times10^{-4}~\mathrm{nm}$, 90\%~\cite{deGouvea:2021uvg})}&\tabincell{c}{$<0.18$ (90\%)\\$<0.22$ (95\%)}\\
\hline
%
%%WPD II&\tabincell{c}{$\alpha\equiv(\sqrt{2}\sigma_\text{rel}/4)^2$\\$\rm[\times10^{-4}]$}&&&\tabincell{c}{$<0.14$  (95\%)}\\
\tabincell{c}{WPD II\\$\alpha~[\times10^{-4}]$}&&\tabincell{c}{$<0.14$  (95\%)}\\
%\cline{2-3}
%\hline
%\tabincell{c}{$\sigma_\text{rel}\equiv2\sqrt{2\alpha}$\\$\rm[\times10^{-2}]$}&&\tabincell{c}{$<1.04$  (95\%)}\\
%% \cline{2-5}
%%&\tabincell{c}{$\sigma_{x}$\\$\rm[nm]$}&&&\tabincell{c}{$>2.63\times10^{-3}$  (90\%)\\$>2.33\times10^{-3}$  (95\%)\\$>1.89\times10^{-3}$  (99\%)}\\%
\hline
%%\alpha\equiv\sigma_{\rm rel}
\tabincell{c}{WPD III\\$\alpha\equiv\sigma_\text{rel}~\rm[\times10^{-2}]$}&\tabincell{c}{\{$<23$ (DYB: $\sigma_\text{rel}<0.23$, 95\%~\cite{An:2016pvi})}\}&\tabincell{c}{$<1.04$ (95\%)}\\%2.38\times10^{-17}<\sigma_{\rm rel}
 \cline{2-3}
\tabincell{c}{$\sigma_{x}\equiv(2\alpha E)^{-1}$\\$\rm[\times 10^{-3}~nm]$}&\tabincell{c}{\{$>10^{-1}$ (DYB: $\sigma_x>10^{-4}~\rm{nm}$, 95\%~\cite{An:2016pvi})}\}&\tabincell{c}{$>2.32$ (95\%)}\\%2.38\times10^{-17}<\sigma_{\rm rel}
% \cline{2-5}
%&\tabincell{c}{$\sigma_{x}\equiv(2\alpha E)^{-1}$\\$\rm[\times 10^{-3}nm]$}&\tabincell{c}{$>10^{-1}$ (DYB: 95\%~\cite{An:2016pvi})}&&\tabincell{c}{$>2.32$ (95\%)}\\%>10^-11cm

\hline
\end{tabular}}
\caption{The limits on the damping parameters for each damping model at JUNO.  The experimental and phenomenological limits in the literature are also shown for comparison.
 }% include experimental limits, phenomenological limits, and our limits for JUNO obtained by this work.
\label{tab:bounds}
\end{table}

Compared to current experimental limits, we find that JUNO will improve the limits on $\tau_3/m_3$ in the $\nu_3$ decay model by a factor of $\sim36$. The limits on $\sigma_{\rm rel}$ (or $\sigma_x$) in the WPD III model can be also improved by a factor of $\sim22$ ($23$). After taking into account the previous limits from phenomenological analysis, we find that JUNO will also impose stronger limits on the damping parameters in WPD I and WPD III. However, the improvement of the bounds on the damping parameters in the QD I, QD II, QD III, $\nu_3$ decay and neutrino absorption scenarios from JUNO is not significant compared to other phenomenological analysis. This is mainly due to the fact that JUNO has a smaller value of $|\Delta m^2_\text{ij}|^{\xi}L^{\beta}/E^\gamma$. From Table~\ref{tab:bounds}, we see that a global joint analysis can be more restrictive in terms of these limits, which provides a promising future direction for JUNO to study these damping effects.

\begin{figure}[h!]
	\centering%{.5\textwidth}
	\includegraphics[width=0.8\textwidth]{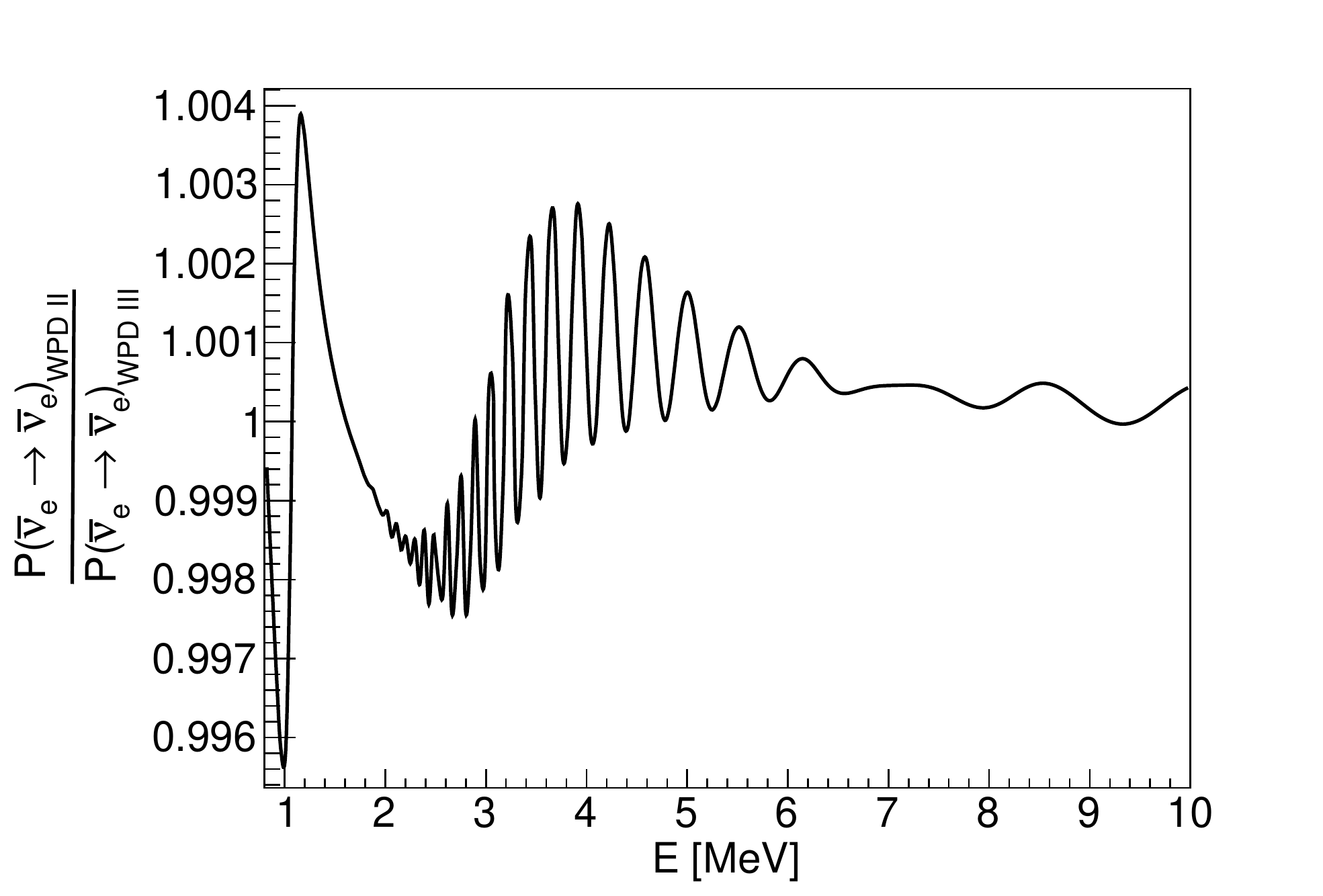}
%\vspace{+.5cm}
\caption{ The ratio of the $\bar{\nu}_e$ survival probabilities between the WPD II and WPD III scenarios as a function of the neutrino energy. Here the oscillation parameters are taken from Table~\ref{tab:pdg2020} and the damping parameter $\sigma_{\rm rel}$ is set to $2.08\times10^{-2}$, which corresponds to a $5\sigma$ CL limit obtained from this work. }%Six years of exposure is assumed.
	\label{fig:possibility_ratio}
\end{figure}

%From Table~\ref{tab:bounds}, 
In the WPD II model, we also replace $\alpha$ with $(\sqrt{2}\sigma_\text{rel}/4)^2$ to study the effect of limit on $\sigma_\text{rel}$ in the absence of the quantum decoherence caused by the dispersion effect. We find that the upper limits on $\sigma_\text{rel}$ for the WPD II and WPD III are about the same, which means that the quantum decoherence caused by the dispersion effect is negligible on the limits on the damping parameters at JUNO. This can be understood from Figure~\ref{fig:possibility_ratio}, which shows that the $\bar{\nu}_e$ survival probabilities described by Eq.~(\ref{eq:wpd1damping-reactor}) and Eq.~(\ref{eq:whole_wp}) are very close at JUNO, and the modification to the $\bar{\nu}_e$ survival probability due to the dispersion effect is less than 0.5\%.

\subsection{Disentangling damping signatures at JUNO}

To compare these eight damping effects, we follow the analysis method described in Ref.~\cite{Blennow:2005yk}. For a fixed set of oscillation parameters and $\alpha$ values in the simulated damping model, we marginalize over the oscillation parameters, $\alpha$ values and all pull parameters in the fitted model. Then, we define a threshold $\alpha_{\rm th}$ as the sensitivity limit for the simulated $\alpha$, i.e., the simulated $\alpha$ must be above this threshold for the simulated damping model to be distinguishable from the fitted model at JUNO. The corresponding sensitivity limits at a 95\% ($3\sigma$) CL obtained through this work are shown in Table~\ref{tab:resu_juno}, where we specifically include the no-damping model among the fitted models. For instance, the QD I model could be distinguished from the no-damping model at the 95\% CL if $ \alpha\gtrsim4.62\times10^{-6}$ $\mathrm{MeV^{2}/m}$.

\begin{table}%\renewcommand\arraystretch{1.5}%\scriptsize%\footnotesize%\small
\renewcommand\arraystretch{1.3}
\resizebox{\textwidth}{!}
{\begin{tabular}{|c|cccccccc|}
\hline
  \tabincell{c}{JUNO\\\tabincell{c}{95\%\\($3\sigma$)}}& \multicolumn{8} {c|}{Simulated damping model}\\
  \hline
\tabincell{c}{Fitted model\\ }& \tabincell{c}{QD I\\$\frac{\alpha}{\mathrm{10^{-6}\frac{MeV^{2}}{m}}}\gtrsim$}&\tabincell{c}{QD II\\$\frac{\alpha}{\mathrm{\frac{10^{-6}}{m}}}\gtrsim$}&\tabincell{c}{QD III\\$\frac{\alpha}{{\mathrm{\frac{10^{-8}}{MeV^{2}\bullet m}}}}\gtrsim$}& \tabincell{c}{Absorption\\$\frac{\alpha}{{\mathrm{\frac{10^{-7}}{MeV\bullet m}}}}\gtrsim$} &\tabincell{c}{$\nu_3$ decay\\$\frac{\alpha}{\mathrm{10^{-4}\frac{MeV}{m}}}\gtrsim$} &\tabincell{c}{WPD I\\$\frac{\alpha}{\mathrm{10^{-3} MeV^{2}}}\gtrsim$} &\tabincell{c}{WPD II\\$\frac{\alpha}{\mathrm{10^{-4}}}\gtrsim$} &\tabincell{c}{WPD III\\$\frac{\alpha}{\mathrm{10^{-2}}}\gtrsim$} \\
\hline
No damping&\tabincell{c}{4.62\\(7.2)}&\tabincell{c}{0.99\\(1.54)}&\tabincell{c}{1.51\\(2.35)}&\tabincell{c}{1.28\\(1.99)}&\tabincell{c}{0.55\\(0.93)}&\tabincell{c}{0.22\\(0.44)}&\tabincell{c}{0.14\\(0.24)}&\tabincell{c}{1.05\\(1.39)}\\
QD I&-&\tabincell{c}{1.05\\(1.62)}&\tabincell{c}{1.51\\(2.35)}&\tabincell{c}{1.28\\(1.99)}&\tabincell{c}{0.55\\(0.93)}&\tabincell{c}{0.22\\(0.44)}&\tabincell{c}{0.14\\(0.24)}&\tabincell{c}{1.05\\(1.39)}\\
QD II&\tabincell{c}{4.82\\(7.5)}&-&\tabincell{c}{1.75\\(2.71)}&\tabincell{c}{1.84\\(2.84)}&\tabincell{c}{0.55\\(0.93)}&\tabincell{c}{0.22\\(0.44)}&\tabincell{c}{0.14\\(0.24)}&\tabincell{c}{1.05\\(1.39)}\\
QD III&\tabincell{c}{4.62\\(7.2)}&\tabincell{c}{1.16\\(1.8)}&-&\tabincell{c}{4.54\\(7.16)}&\tabincell{c}{0.55\\(0.93)}&\tabincell{c}{0.22\\(0.44)}&\tabincell{c}{0.14\\(0.24)}&\tabincell{c}{1.05\\(1.39)}\\
Absorption&\tabincell{c}{4.62\\(7.2)}&\tabincell{c}{1.43\\(2.24)}&\tabincell{c}{5.26\\(8.21)}&-&\tabincell{c}{0.55\\(0.93)}&\tabincell{c}{0.22\\(0.44)}&\tabincell{c}{0.14\\(0.24)}&\tabincell{c}{1.05\\(1.39)}\\
$\nu_{3}$ decay&\tabincell{c}{4.62\\(7.2)}&\tabincell{c}{0.99\\(1.54)}&\tabincell{c}{1.51\\(2.35)}&\tabincell{c}{1.28\\(1.99)}&-&\tabincell{c}{4.03\\(7.17)}&\tabincell{c}{10.48\\(16.64)}&\tabincell{c}{8.88\\(11.04)}\\
WPD I&\tabincell{c}{4.62\\(7.2)}&\tabincell{c}{0.99\\(1.54)}&\tabincell{c}{1.51\\(2.35)}&\tabincell{c}{1.28\\(1.99)}&\tabincell{c}{4.4\\(-)}&-&\tabincell{c}{3.2\\(39.2)}&\tabincell{c}{4.72\\(15.68)}\\
WPD II&\tabincell{c}{4.62\\(7.2)}&\tabincell{c}{0.99\\(1.54)}&\tabincell{c}{1.51\\(2.35)}&\tabincell{c}{1.28\\(1.99)}&-&\tabincell{c}{10\\(17.2)}&-&\tabincell{c}{21.76\\(25.04)}\\
WPD III&\tabincell{c}{4.62\\(7.2)}&\tabincell{c}{0.99\\(1.54)}&\tabincell{c}{1.51\\(2.35)}&\tabincell{c}{1.28\\(1.99)}&-&\tabincell{c}{9.12\\(15.68)}&\tabincell{c}{66.8\\(88.4)}&-\\
\hline
\end{tabular}
}
\caption{The sensitivity limits on $\alpha$ for which a certain simulated damping model (in columns) could be distinguished from a certain fitted model (in rows) at JUNO.  %Six years of exposure is assumed.
}  
\label{tab:resu_juno}
\end{table}

%The values with no damping in Table~\ref{tab:resu_juno} represent the upper bound of the simulated $\alpha$ required to identify each damping effect model.  compensated, scenario
In the rows representing $\nu_3$ decay versus WPD-like models, there are no corresponding $\alpha_\text{th}$ values at the $3\sigma$ CL since the $\chi^2$ are below 6.4 for all $\alpha$ values in the simulated $\nu_3$ decay model. This can be attributed to the distortion of the standard $\bar{\nu}_e$ survival probability spectrum caused by the $\nu_3$ decay, which can be easily compensated for by shifting the neutrino oscillation parameters and $\alpha$ in the fitted WPD-like models. In the columns representing WPD-like models, the values with other WPD-like or $\nu_3$ decay scenarios are several orders of magnitude greater than the values with QD-like models. Thus, if a WPD-like model exists in nature, it will be much more difficult to distinguish it from other WPD-like scenarios or from a $\nu_3$ decay scenario as compared to a QD-like model. 
%However, the values in the columns representing QD-like effects indicate that the difficulty of distinguishing one of them from the remaining models is roughly the same. The reason for the difference in the difficulty of distinguishing these categories of damping effect models from each other is the difference in the detailed smearing behavior described in Section~\ref{subsec:classification}.

%%%%%%%%%%%%%%%%%%%%%%%%%%%%%%%% 
\section{Conclusions}
\label{sec:conclusions}

In this paper, we systematically study the phenomenology of damping signatures at JUNO, a medium-baseline reactor neutrino oscillation experiment. As the benchmark models in this work, we analyze several new physics scenarios, including quantum decoherence, $\nu_3$ decay, neutrino absorption, and wave packet decoherence. Based on a six-year exposure and five main background sources for the JUNO detector, we demonstrate how to test and disentangle the fine-scale spectral structure caused by the damping effects. The exclusion sensitivities on the damping parameters at JUNO for each benchmark model are listed in Table~\ref{tab:bounds}. Compared to current experimental limits, JUNO will significantly improve the limits on $\tau_3/m_3$ in the $\nu_3$ decay model, the width of the neutrino wave packet $\sigma_x$, and the intrinsic relative dispersion of neutrino momentum $\sigma_{\rm rel}$ by a factor of $\sim36$, $23$ and $22$, respectively. Furthermore, we find that the quantum decoherence caused by the dispersion effect is negligible at JUNO. Finally, we find that compared to the QD-like models, the WPD-like and $\nu_3$ decay models are much more difficult to distinguish from each other at JUNO.

\acknowledgments
We are grateful for the ongoing cooperation from the China General Nuclear Power Group.
This work was supported by
the Chinese Academy of Sciences,
the National Key R\&D Program of China,
the CAS Center for Excellence in Particle Physics,
Wuyi University,
and the Tsung-Dao Lee Institute of Shanghai Jiao Tong University in China,
the Institut National de Physique Nucl\'eaire et de Physique des Particules (IN2P3) in France,
the Istituto Nazionale di Fisica Nucleare (INFN) in Italy,
the Italian-Chinese collaborative research program MAECI-NSFC,
the Fond de la Recherche Scientifique (F.R.S-FNRS) and FWO under the ``Excellence of Science – EOS” in Belgium,
the Conselho Nacional de Desenvolvimento Cient\'ifico e Tecnol\`ogico in Brazil,
the Agencia Nacional de Investigacion y Desarrollo in Chile,
the Charles University Research Centre and the Ministry of Education, Youth, and Sports in Czech Republic,
the Deutsche Forschungsgemeinschaft (DFG), the Helmholtz Association, and the Cluster of Excellence PRISMA+ in Germany,
the Joint Institute of Nuclear Research (JINR) and Lomonosov Moscow State University in Russia,
the joint Russian Science Foundation (RSF) and National Natural Science Foundation of China (NSFC) research program,
the MOST and MOE in Taiwan,
the Chulalongkorn University and Suranaree University of Technology in Thailand,
and the University of California at Irvine in USA.

\bibliographystyle{JHEP}
\bibliography{ref}

\providecommand{\href}[2]{#2}\begingroup\raggedright\begin{thebibliography}{10}

\bibitem{Pontecorvo:1957qd}
B.~Pontecorvo, {\it {Inverse beta processes and nonconservation of lepton
  charge}},  {\em Zh. Eksp. Teor. Fiz.} {\bf 34} (1957) 247. [Sov. Phys. JETP,
  7 (1958), pp. 172-173].

\bibitem{Super-Kamiokande:1998kpq}
{\bf Super-Kamiokande} Collaboration, Y.~Fukuda et~al., {\it {Evidence for
  oscillation of atmospheric neutrinos}},  {\em Phys. Rev. Lett.} {\bf 81}
  (1998) 1562--1567, [\href{http://arxiv.org/abs/hep-ex/9807003}{{\tt
  hep-ex/9807003}}].

\bibitem{Giganti:2017fhf}
C.~Giganti, S.~Lavignac, and M.~Zito, {\it {Neutrino oscillations: The rise of
  the PMNS paradigm}},  {\em Prog. Part. Nucl. Phys.} {\bf 98} (2018) 1--54,
  [\href{http://arxiv.org/abs/1710.00715}{{\tt arXiv:1710.00715}}].

\bibitem{Zyla:2020zbs}
{\bf Particle Data Group} Collaboration, P.~A. Zyla et~al., {\it {Review of
  Particle Physics}},  {\em PTEP} {\bf 2020} (2020), no.~8 083C01.

\bibitem{NOvA:2017ohq}
{\bf NOvA} Collaboration, P.~Adamson et~al., {\it {Measurement of the neutrino
  mixing angle $\theta_{23}$ in NOvA}},  {\em Phys. Rev. Lett.} {\bf 118}
  (2017), no.~15 151802, [\href{http://arxiv.org/abs/1701.05891}{{\tt
  arXiv:1701.05891}}].

\bibitem{Super-Kamiokande:2017yvm}
{\bf Super-Kamiokande} Collaboration, K.~Abe et~al., {\it {Atmospheric neutrino
  oscillation analysis with external constraints in Super-Kamiokande I-IV}},
  {\em Phys. Rev. D} {\bf 97} (2018), no.~7 072001,
  [\href{http://arxiv.org/abs/1710.09126}{{\tt arXiv:1710.09126}}].

\bibitem{T2K:2019bcf}
{\bf T2K} Collaboration, K.~Abe et~al., {\it {Constraint on the
  matter\textendash{}antimatter symmetry-violating phase in neutrino
  oscillations}},  {\em Nature} {\bf 580} (2020), no.~7803 339--344,
  [\href{http://arxiv.org/abs/1910.03887}{{\tt arXiv:1910.03887}}]. [Erratum:
  Nature 583, E16 (2020)].

\bibitem{Acciarri:2015uup}
{\bf DUNE} Collaboration, R.~Acciarri et~al., {\it {Long-Baseline Neutrino
  Facility (LBNF) and Deep Underground Neutrino Experiment (DUNE)}: {Conceptual
  Design Report, Volume 2: The Physics Program for DUNE at LBNF}},
  \href{http://arxiv.org/abs/1512.06148}{{\tt arXiv:1512.06148}}.

\bibitem{DUNE:2020jqi}
{\bf DUNE} Collaboration, B.~Abi et~al., {\it {Long-baseline neutrino
  oscillation physics potential of the DUNE experiment}},  {\em Eur. Phys. J.
  C} {\bf 80} (2020), no.~10 978, [\href{http://arxiv.org/abs/2006.16043}{{\tt
  arXiv:2006.16043}}].

\bibitem{Abe:2018uyc}
{\bf Hyper-Kamiokande} Collaboration, K.~Abe et~al., {\it {Hyper-Kamiokande
  Design Report}},  \href{http://arxiv.org/abs/1805.04163}{{\tt
  arXiv:1805.04163}}.

\bibitem{An:2015jdp}
{\bf JUNO} Collaboration, F.~An et~al., {\it {Neutrino Physics with JUNO}},
  {\em J. Phys. G} {\bf 43} (2016), no.~3 030401,
  [\href{http://arxiv.org/abs/1507.05613}{{\tt arXiv:1507.05613}}].

\bibitem{Abusleme:2021zrw}
{\bf JUNO} Collaboration, A.~Abusleme et~al., {\it {JUNO Physics and
  Detector}},  \href{http://arxiv.org/abs/2104.02565}{{\tt arXiv:2104.02565}}.

\bibitem{Blennow:2005yk}
M.~Blennow, T.~Ohlsson, and W.~Winter, {\it {Damping signatures in future
  neutrino oscillation experiments}},  {\em JHEP} {\bf 06} (2005) 049,
  [\href{http://arxiv.org/abs/hep-ph/0502147}{{\tt hep-ph/0502147}}].

\bibitem{Blennow:2006hd}
M.~Blennow, {\it {Damping signatures in future neutrino oscillation
  experiments}},  {\em Nucl. Phys. B Proc. Suppl.} {\bf 155} (2006) 195--196.

\bibitem{MINOS:2010fgd}
{\bf MINOS} Collaboration, P.~Adamson et~al., {\it {Search for sterile neutrino
  mixing in the MINOS long baseline experiment}},  {\em Phys. Rev. D} {\bf 81}
  (2010) 052004, [\href{http://arxiv.org/abs/1001.0336}{{\tt
  arXiv:1001.0336}}].

\bibitem{Abrahao:2015rba}
T.~Abrah\~ao, H.~Minakata, H.~Nunokawa, and A.~A. Quiroga, {\it {Constraint on
  Neutrino Decay with Medium-Baseline Reactor Neutrino Oscillation
  Experiments}},  {\em JHEP} {\bf 11} (2015) 001,
  [\href{http://arxiv.org/abs/1506.02314}{{\tt arXiv:1506.02314}}].

\bibitem{Choubey:2018cfz}
S.~Choubey, D.~Dutta, and D.~Pramanik, {\it {Invisible neutrino decay in the
  light of NOvA and T2K data}},  {\em JHEP} {\bf 08} (2018) 141,
  [\href{http://arxiv.org/abs/1805.01848}{{\tt arXiv:1805.01848}}].

\bibitem{Porto-Silva:2020gma}
Y.~P. Porto-Silva, S.~Prakash, O.~L.~G. Peres, H.~Nunokawa, and H.~Minakata,
  {\it {Constraining visible neutrino decay at KamLAND and JUNO}},  {\em Eur.
  Phys. J. C} {\bf 80} (2020), no.~10 999,
  [\href{http://arxiv.org/abs/2002.12134}{{\tt arXiv:2002.12134}}].

\bibitem{Ghoshal:2020hyo}
A.~Ghoshal, A.~Giarnetti, and D.~Meloni, {\it {Neutrino Invisible Decay at
  DUNE: a multi-channel analysis}},  {\em J. Phys. G} {\bf 48} (2021), no.~5
  055004, [\href{http://arxiv.org/abs/2003.09012}{{\tt arXiv:2003.09012}}].

\bibitem{Fogli:2007tx}
G.~L. Fogli, E.~Lisi, A.~Marrone, D.~Montanino, and A.~Palazzo, {\it {Probing
  non-standard decoherence effects with solar and KamLAND neutrinos}},  {\em
  Phys. Rev. D} {\bf 76} (2007) 033006,
  [\href{http://arxiv.org/abs/0704.2568}{{\tt arXiv:0704.2568}}].

\bibitem{Chan:2015mca}
Y.-L. Chan, M.~C. Chu, K.~M. Tsui, C.~F. Wong, and J.~Xu, {\it {Wave-packet
  treatment of reactor neutrino oscillation experiments and its implications on
  determining the neutrino mass hierarchy}},  {\em Eur. Phys. J. C} {\bf 76}
  (2016), no.~6 310, [\href{http://arxiv.org/abs/1507.06421}{{\tt
  arXiv:1507.06421}}].

\bibitem{An:2016pvi}
{\bf Daya Bay} Collaboration, F.~P. An et~al., {\it {Study of the wave packet
  treatment of neutrino oscillation at Daya Bay}},  {\em Eur. Phys. J. C} {\bf
  77} (2017), no.~9 606, [\href{http://arxiv.org/abs/1608.01661}{{\tt
  arXiv:1608.01661}}].

\bibitem{Coelho:2017zes}
J.~A.~B. Coelho, W.~A. Mann, and S.~S. Bashar, {\it {Nonmaximal $\theta_{23}$
  mixing at NOvA from neutrino decoherence}},  {\em Phys. Rev. Lett.} {\bf 118}
  (2017), no.~22 221801, [\href{http://arxiv.org/abs/1702.04738}{{\tt
  arXiv:1702.04738}}].

\bibitem{deGouvea:2020hfl}
A.~de~Gouvea, V.~de~Romeri, and C.~A. Ternes, {\it {Probing neutrino quantum
  decoherence at reactor experiments}},  {\em JHEP} {\bf 08} (2020) 018,
  [\href{http://arxiv.org/abs/2005.03022}{{\tt arXiv:2005.03022}}].

\bibitem{Cheng:2020jje}
Z.~Cheng, W.~Wang, C.~F. Wong, and J.~Zhang, {\it {Studying the neutrino
  wave-packet effects at medium-baseline reactor neutrino oscillation
  experiments and the potential benefits of an extra detector}},  {\em Nucl.
  Phys. B} {\bf 964} (2021) 115304,
  [\href{http://arxiv.org/abs/2009.06450}{{\tt arXiv:2009.06450}}].

\bibitem{Liu:1997zd}
Y.~Liu, J.-L. Chen, and M.-L. Ge, {\it {A Constraint on EHNS parameters from
  solar neutrino problem}},  {\em J. Phys. G} {\bf 24} (1998) 2289--2296,
  [\href{http://arxiv.org/abs/hep-ph/9711381}{{\tt hep-ph/9711381}}].

\bibitem{Shafaq:2021lju}
S.~Shafaq, T.~Kushwaha, and P.~Mehta, {\it {Investigating Leggett-Garg
  inequality in neutrino oscillations -- role of decoherence and decay}},
  \href{http://arxiv.org/abs/2112.12726}{{\tt arXiv:2112.12726}}.

\bibitem{Ribeiro:2007jq}
N.~C. Ribeiro, H.~Nunokawa, T.~Kajita, S.~Nakayama, P.~Ko, and H.~Minakata,
  {\it {Probing Nonstandard Neutrino Physics by Two Identical Detectors with
  Different Baselines}},  {\em Phys. Rev. D} {\bf 77} (2008) 073007,
  [\href{http://arxiv.org/abs/0712.4314}{{\tt arXiv:0712.4314}}].

\bibitem{Farzan:2008zv}
Y.~Farzan, T.~Schwetz, and A.~Y. Smirnov, {\it {Reconciling results of LSND,
  MiniBooNE and other experiments with soft decoherence}},  {\em JHEP} {\bf 07}
  (2008) 067, [\href{http://arxiv.org/abs/0805.2098}{{\tt arXiv:0805.2098}}].

\bibitem{Machado:2011tn}
P.~A.~N. Machado, H.~Nunokawa, F.~A. Pereira~dos Santos, and
  R.~Zukanovich~Funchal, {\it {Testing Nonstandard Neutrino Properties with a
  M\"{o}ssbauer Oscillation Experiment}},  {\em JHEP} {\bf 11} (2011) 136,
  [\href{http://arxiv.org/abs/1108.3339}{{\tt arXiv:1108.3339}}].

\bibitem{deOliveira:2013dia}
R.~L.~N. de~Oliveira, M.~M. Guzzo, and P.~C. de~Holanda, {\it {Quantum
  Dissipation and $C\!P$ Violation in MINOS}},  {\em Phys. Rev. D} {\bf 89}
  (2014), no.~5 053002, [\href{http://arxiv.org/abs/1401.0033}{{\tt
  arXiv:1401.0033}}].

\bibitem{Coloma:2018idr}
P.~Coloma, J.~Lopez-Pavon, I.~Martinez-Soler, and H.~Nunokawa, {\it
  {Decoherence in Neutrino Propagation Through Matter, and Bounds from
  IceCube/DeepCore}},  {\em Eur. Phys. J. C} {\bf 78} (2018), no.~8 614,
  [\href{http://arxiv.org/abs/1803.04438}{{\tt arXiv:1803.04438}}].

\bibitem{Gomes:2020muc}
A.~L.~G. Gomes, R.~A. Gomes, and O.~L.~G. Peres, {\it {Quantum decoherence and
  relaxation in neutrinos using long-baseline data}},
  \href{http://arxiv.org/abs/2001.09250}{{\tt arXiv:2001.09250}}.

\bibitem{Liu:1997km}
Y.~Liu, L.-z. Hu, and M.-L. Ge, {\it {The Effect of quantum mechanics violation
  on neutrino oscillation}},  {\em Phys. Rev. D} {\bf 56} (1997) 6648--6652.

\bibitem{Lisi:2000zt}
E.~Lisi, A.~Marrone, and D.~Montanino, {\it {Probing possible decoherence
  effects in atmospheric neutrino oscillations}},  {\em Phys. Rev. Lett.} {\bf
  85} (2000) 1166--1169, [\href{http://arxiv.org/abs/hep-ph/0002053}{{\tt
  hep-ph/0002053}}].

\bibitem{Morgan:2004vv}
D.~Morgan, E.~Winstanley, J.~Brunner, and L.~F. Thompson, {\it {Probing quantum
  decoherence in atmospheric neutrino oscillations with a neutrino telescope}},
   {\em Astropart. Phys.} {\bf 25} (2006) 311--327,
  [\href{http://arxiv.org/abs/astro-ph/0412618}{{\tt astro-ph/0412618}}].

\bibitem{Mavromatos:2006yy}
N.~E. Mavromatos and S.~Sarkar, {\it {Methods of approaching decoherence in the
  flavour sector due to space-time foam}},  {\em Phys. Rev. D} {\bf 74} (2006)
  036007, [\href{http://arxiv.org/abs/hep-ph/0606048}{{\tt hep-ph/0606048}}].

\bibitem{Mehta:2011qb}
P.~Mehta and W.~Winter, {\it {Interplay of energy dependent astrophysical
  neutrino flavor ratios and new physics effects}},  {\em JCAP} {\bf 03} (2011)
  041, [\href{http://arxiv.org/abs/1101.2673}{{\tt arXiv:1101.2673}}].

\bibitem{Bakhti:2015dca}
P.~Bakhti, Y.~Farzan, and T.~Schwetz, {\it {Revisiting the quantum decoherence
  scenario as an explanation for the LSND anomaly}},  {\em JHEP} {\bf 05}
  (2015) 007, [\href{http://arxiv.org/abs/1503.05374}{{\tt arXiv:1503.05374}}].

\bibitem{Gomes:2016ixi}
G.~Balieiro~Gomes, M.~M. Guzzo, P.~C. de~Holanda, and R.~L.~N. Oliveira, {\it
  {Parameter Limits for Neutrino Oscillation with Decoherence in KamLAND}},
  {\em Phys. Rev. D} {\bf 95} (2017), no.~11 113005,
  [\href{http://arxiv.org/abs/1603.04126}{{\tt arXiv:1603.04126}}].

\bibitem{Gonzalez-Garcia:2008mgl}
M.~C. Gonzalez-Garcia and M.~Maltoni, {\it {Status of Oscillation plus Decay of
  Atmospheric and Long-Baseline Neutrinos}},  {\em Phys. Lett. B} {\bf 663}
  (2008) 405--409, [\href{http://arxiv.org/abs/0802.3699}{{\tt
  arXiv:0802.3699}}].

\bibitem{Gomes:2014yua}
R.~A. Gomes, A.~L.~G. Gomes, and O.~L.~G. Peres, {\it {Constraints on neutrino
  decay lifetime using long-baseline charged and neutral current data}},  {\em
  Phys. Lett. B} {\bf 740} (2015) 345--352,
  [\href{http://arxiv.org/abs/1407.5640}{{\tt arXiv:1407.5640}}].

\bibitem{Pagliaroli:2016zab}
G.~Pagliaroli, N.~Di~Marco, and M.~Mannarelli, {\it {Enhanced tau neutrino
  appearance through invisible decay}},  {\em Phys. Rev. D} {\bf 93} (2016),
  no.~11 113011, [\href{http://arxiv.org/abs/1603.08696}{{\tt
  arXiv:1603.08696}}].

\bibitem{Giunti:1991sx}
C.~Giunti, C.~W. Kim, and U.~W. Lee, {\it {Coherence of neutrino oscillations
  in vacuum and matter in the wave packet treatment}},  {\em Phys. Lett. B}
  {\bf 274} (1992) 87--94.

\bibitem{Giunti:1997wq}
C.~Giunti and C.~W. Kim, {\it {Coherence of neutrino oscillations in the wave
  packet approach}},  {\em Phys. Rev. D} {\bf 58} (1998) 017301,
  [\href{http://arxiv.org/abs/hep-ph/9711363}{{\tt hep-ph/9711363}}].

\bibitem{Giunti:2003ax}
C.~Giunti, {\it {Coherence and wave packets in neutrino oscillations}},  {\em
  Found. Phys. Lett.} {\bf 17} (2004) 103--124,
  [\href{http://arxiv.org/abs/hep-ph/0302026}{{\tt hep-ph/0302026}}].

\bibitem{Blasone:2015lya}
M.~Blasone, F.~Dell'Anno, S.~De~Siena, and F.~Illuminati, {\it {Flavor
  entanglement in neutrino oscillations in the wave packet description}},  {\em
  EPL} {\bf 112} (2015), no.~2 20007,
  [\href{http://arxiv.org/abs/1510.06761}{{\tt arXiv:1510.06761}}].

\bibitem{Kersten:2015kio}
J.~Kersten and A.~Y. Smirnov, {\it {Decoherence and oscillations of supernova
  neutrinos}},  {\em Eur. Phys. J. C} {\bf 76} (2016), no.~6 339,
  [\href{http://arxiv.org/abs/1512.09068}{{\tt arXiv:1512.09068}}].

\bibitem{deGouvea:2021uvg}
A.~de~Gouv\^ea, V.~De~Romeri, and C.~A. Ternes, {\it {Combined analysis of
  neutrino decoherence at reactor experiments}},  {\em JHEP} {\bf 06} (2021)
  042, [\href{http://arxiv.org/abs/2104.05806}{{\tt arXiv:2104.05806}}].

\bibitem{Ohlsson:2000mj}
T.~Ohlsson, {\it {Equivalence between neutrino oscillations and neutrino
  decoherence}},  {\em Phys. Lett. B} {\bf 502} (2001) 159--166,
  [\href{http://arxiv.org/abs/hep-ph/0012272}{{\tt hep-ph/0012272}}].

\bibitem{Naumov:2010um}
D.~V. Naumov and V.~A. Naumov, {\it {A Diagrammatic treatment of neutrino
  oscillations}},  {\em J. Phys. G} {\bf 37} (2010) 105014,
  [\href{http://arxiv.org/abs/1008.0306}{{\tt arXiv:1008.0306}}].

\bibitem{Naumov:2020yyv}
D.~V. Naumov and V.~A. Naumov, {\it {Quantum Field Theory of Neutrino
  Oscillations}},  {\em Phys. Part. Nucl.} {\bf 51} (2020), no.~1 1--106.

\bibitem{Ellis:1996bz}
J.~R. Ellis, N.~E. Mavromatos, D.~V. Nanopoulos, and E.~Winstanley, {\it
  {Quantum decoherence in a four-dimensional black hole background}},  {\em
  Mod. Phys. Lett. A} {\bf 12} (1997) 243--256,
  [\href{http://arxiv.org/abs/gr-qc/9602011}{{\tt gr-qc/9602011}}].

\bibitem{Ellis:1997jw}
J.~R. Ellis, N.~E. Mavromatos, and D.~V. Nanopoulos, {\it {Quantum decoherence
  in a D foam background}},  {\em Mod. Phys. Lett. A} {\bf 12} (1997)
  1759--1773, [\href{http://arxiv.org/abs/hep-th/9704169}{{\tt
  hep-th/9704169}}].

\bibitem{Lindner:2001fx}
M.~Lindner, T.~Ohlsson, and W.~Winter, {\it {A Combined treatment of neutrino
  decay and neutrino oscillations}},  {\em Nucl. Phys. B} {\bf 607} (2001)
  326--354, [\href{http://arxiv.org/abs/hep-ph/0103170}{{\tt hep-ph/0103170}}].

\bibitem{Joshipura:2002fb}
A.~S. Joshipura, E.~Masso, and S.~Mohanty, {\it {Constraints on decay plus
  oscillation solutions of the solar neutrino problem}},  {\em Phys. Rev. D}
  {\bf 66} (2002) 113008, [\href{http://arxiv.org/abs/hep-ph/0203181}{{\tt
  hep-ph/0203181}}].

\bibitem{Beacom:2002cb}
J.~F. Beacom and N.~F. Bell, {\it {Do Solar Neutrinos Decay?}},  {\em Phys.
  Rev. D} {\bf 65} (2002) 113009,
  [\href{http://arxiv.org/abs/hep-ph/0204111}{{\tt hep-ph/0204111}}].

\bibitem{Fogli:2003th}
G.~L. Fogli, E.~Lisi, A.~Marrone, and D.~Montanino, {\it {Status of atmospheric
  nu(mu) ---\ensuremath{>} nu(tau) oscillations and decoherence after the first
  K2K spectral data}},  {\em Phys. Rev. D} {\bf 67} (2003) 093006,
  [\href{http://arxiv.org/abs/hep-ph/0303064}{{\tt hep-ph/0303064}}].

\bibitem{Baerwald:2012kc}
P.~Baerwald, M.~Bustamante, and W.~Winter, {\it {Neutrino Decays over
  Cosmological Distances and the Implications for Neutrino Telescopes}},  {\em
  JCAP} {\bf 10} (2012) 020, [\href{http://arxiv.org/abs/1208.4600}{{\tt
  arXiv:1208.4600}}].

\bibitem{Picoreti:2015ika}
R.~Picoreti, M.~M. Guzzo, P.~C. de~Holanda, and O.~L.~G. Peres, {\it {Neutrino
  Decay and Solar Neutrino Seasonal Effect}},  {\em Phys. Lett. B} {\bf 761}
  (2016) 70--73, [\href{http://arxiv.org/abs/1506.08158}{{\tt
  arXiv:1506.08158}}].

\bibitem{Gago:2017zzy}
A.~M. Gago, R.~A. Gomes, A.~L.~G. Gomes, J.~Jones-Perez, and O.~L.~G. Peres,
  {\it {Visible neutrino decay in the light of appearance and disappearance
  long baseline experiments}},  {\em JHEP} {\bf 11} (2017) 022,
  [\href{http://arxiv.org/abs/1705.03074}{{\tt arXiv:1705.03074}}].

\bibitem{SNO:2018pvg}
{\bf SNO} Collaboration, B.~Aharmim et~al., {\it {Constraints on Neutrino
  Lifetime from the Sudbury Neutrino Observatory}},  {\em Phys. Rev. D} {\bf
  99} (2019), no.~3 032013, [\href{http://arxiv.org/abs/1812.01088}{{\tt
  arXiv:1812.01088}}].

\bibitem{Akhmedov:2009rb}
E.~K. Akhmedov and A.~Y. Smirnov, {\it {Paradoxes of neutrino oscillations}},
  {\em Phys. Atom. Nucl.} {\bf 72} (2009) 1363--1381,
  [\href{http://arxiv.org/abs/0905.1903}{{\tt arXiv:0905.1903}}].

\bibitem{Adler:2000vfa}
S.~L. Adler, {\it {Comment on a proposed Super-Kamiokande test for quantum
  gravity induced decoherence effects}},  {\em Phys. Rev. D} {\bf 62} (2000)
  117901, [\href{http://arxiv.org/abs/hep-ph/0005220}{{\tt hep-ph/0005220}}].

\bibitem{Wolfenstein:1977ue}
L.~Wolfenstein, {\it {Neutrino Oscillations in Matter}},  {\em Phys. Rev. D}
  {\bf 17} (1978) 2369--2374.

\bibitem{Mikheyev:1985zog}
S.~P. Mikheyev and A.~Y. Smirnov, {\it {Resonance Amplification of Oscillations
  in Matter and Spectroscopy of Solar Neutrinos}},  {\em Sov. J. Nucl. Phys.}
  {\bf 42} (1985) 913--917.

\bibitem{Li:2016txk}
Y.-F. Li, Y.~Wang, and Z.-z. Xing, {\it {Terrestrial matter effects on reactor
  antineutrino oscillations at JUNO or RENO-50: how small is small?}},  {\em
  Chin. Phys. C} {\bf 40} (2016), no.~9 091001,
  [\href{http://arxiv.org/abs/1605.00900}{{\tt arXiv:1605.00900}}].

\bibitem{Khan:2019doq}
A.~N. Khan, H.~Nunokawa, and S.~J. Parke, {\it {Why matter effects matter for
  JUNO}},  {\em Phys. Lett. B} {\bf 803} (2020) 135354,
  [\href{http://arxiv.org/abs/1910.12900}{{\tt arXiv:1910.12900}}].

\bibitem{Ge:2012wj}
S.-F. Ge, K.~Hagiwara, N.~Okamura, and Y.~Takaesu, {\it {Determination of mass
  hierarchy with medium baseline reactor neutrino experiments}},  {\em JHEP}
  {\bf 05} (2013) 131, [\href{http://arxiv.org/abs/1210.8141}{{\tt
  arXiv:1210.8141}}].

\bibitem{DayaBay:2016ssb}
{\bf Daya Bay} Collaboration, F.~P. An et~al., {\it {Improved Measurement of
  the Reactor Antineutrino Flux and Spectrum at Daya Bay}},  {\em Chin. Phys.
  C} {\bf 41} (2017), no.~1 013002,
  [\href{http://arxiv.org/abs/1607.05378}{{\tt arXiv:1607.05378}}].

\bibitem{Huber:2011wv}
P.~Huber, {\it {On the determination of anti-neutrino spectra from nuclear
  reactors}},  {\em Phys. Rev. C} {\bf 84} (2011) 024617,
  [\href{http://arxiv.org/abs/1106.0687}{{\tt arXiv:1106.0687}}]. [Erratum:
  Phys.Rev.C 85, 029901 (2012)].

\bibitem{Mueller:2011nm}
T.~A. Mueller et~al., {\it {Improved Predictions of Reactor Antineutrino
  Spectra}},  {\em Phys. Rev. C} {\bf 83} (2011) 054615,
  [\href{http://arxiv.org/abs/1101.2663}{{\tt arXiv:1101.2663}}].

\bibitem{Strumia:2003zx}
A.~Strumia and F.~Vissani, {\it {Precise quasielastic neutrino/nucleon
  cross-section}},  {\em Phys. Lett. B} {\bf 564} (2003) 42--54,
  [\href{http://arxiv.org/abs/astro-ph/0302055}{{\tt astro-ph/0302055}}].

\bibitem{Dighe:2003be}
A.~S. Dighe, M.~T. Keil, and G.~G. Raffelt, {\it {Detecting the neutrino mass
  hierarchy with a supernova at IceCube}},  {\em JCAP} {\bf 06} (2003) 005,
  [\href{http://arxiv.org/abs/hep-ph/0303210}{{\tt hep-ph/0303210}}].

\bibitem{Abusleme:2020lur}
{\bf JUNO} Collaboration, A.~Abusleme et~al., {\it {Calibration Strategy of the
  JUNO Experiment}},  {\em JHEP} {\bf 03} (2021) 004,
  [\href{http://arxiv.org/abs/2011.06405}{{\tt arXiv:2011.06405}}].

\bibitem{Capozzi:2013psa}
F.~Capozzi, E.~Lisi, and A.~Marrone, {\it {Neutrino mass hierarchy and electron
  neutrino oscillation parameters with one hundred thousand reactor events}},
  {\em Phys. Rev. D} {\bf 89} (2014), no.~1 013001,
  [\href{http://arxiv.org/abs/1309.1638}{{\tt arXiv:1309.1638}}].

\bibitem{Wang:2016vua}
H.~Wang, L.~Zhan, Y.-F. Li, G.~Cao, and S.~Chen, {\it {Mass hierarchy
  sensitivity of medium baseline reactor neutrino experiments with multiple
  detectors}},  {\em Nucl. Phys. B} {\bf 918} (2017) 245--256,
  [\href{http://arxiv.org/abs/1602.04442}{{\tt arXiv:1602.04442}}].

\end{thebibliography}\endgroup

\end{document}